\pdfoutput=1
\documentclass[12pt]{article}

\newcommand{\newc}{\newcommand}
\newc{\beq}{\begin{equation}}
\newc{\eeq}{\end{equation}}
\newc{\bal}{\begin{align}}
\newc{\eal}{\end{align}}
\newc{\ba}{\begin{eqnarray}}
\newc{\ea}{\end{eqnarray}}
\newc{\bea}{\begin{eqnarray*}}
\newc{\eea}{\end{eqnarray*}}
\newc{\alp}{\alpha}
\newc{\eps}{\epsilon}
\newc{\vph}{\varphi}
\newc{\vhp}{\varphi}
\newc{\tx}{\tilde{x}}
\newcommand{\comm}[1]{}

\def\rf{\eqref}
\def\Scz{Schwarz\-schild}
\def\RN{Reiss\-ner-Nord\-sr\"om}
\def\Jl#1#2{{\it #1} {\bf #2}\ }
\def\CQG#1 {\Jl{Class. Quantum Grav.}{#1}}
\def\GC#1 {\Jl{Grav. Cosmol.}{#1}}
\def\GRG#1 {\Jl{Gen. Rel. Grav.}{#1}}
\def\PRD#1 {\Jl{Phys. Rev. D}{#1}}
\def\PRL#1 {\Jl{Phys. Rev. Lett.}{#1}}
\usepackage{graphicx,rotating,colortbl}
\usepackage{enumerate}
\usepackage{jcappub}
\usepackage{amsmath}
\usepackage{amssymb}
\usepackage{braket}
\usepackage{graphicx}
\usepackage{hyperref}
\usepackage{mathrsfs}
\usepackage{subcaption}
\usepackage{empheq}
\usepackage{appendix}
\usepackage{natbib}
\usepackage[dvipsnames]{xcolor}
\usepackage{float}\usepackage[top=1in, bottom=0.4in, left=0.8in, right=0.8in, includefoot]{geometry} 
\usepackage[autostyle]{csquotes}
\usepackage{color,soul}
\usepackage{mathtools}%\coloneqq

\definecolor{DarkViolet}{RGB}{148,0,211}
\definecolor{DarkBlue}{RGB}{0,0,153}

\hypersetup{colorlinks,bookmarksopen,
bookmarksnumbered,
citecolor={DarkViolet},
linkcolor={DarkViolet},
urlcolor={DarkViolet},
pdfstartview=FitH} 

\usepackage{eulervm}
\usepackage{scalerel}
\usepackage{tikz}
\usetikzlibrary{svg.path}
\definecolor{orcidlogocol}{HTML}{A6CE39}
\tikzset{
  orcidlogo/.pic={
    \fill[orcidlogocol] svg{M256,128c0,70.7-57.3,128-128,128C57.3,256,0,198.7,0,128C0,57.3,57.3,0,128,0C198.7,0,256,57.3,256,128z};
    \fill[white] svg{M86.3,186.2H70.9V79.1h15.4v48.4V186.2z}
                 svg{M108.9,79.1h41.6c39.6,0,57,28.3,57,53.6c0,27.5-21.5,53.6-56.8,53.6h-41.8V79.1z M124.3,172.4h24.5c34.9,0,42.9-26.5,42.9-39.7c0-21.5-13.7-39.7-43.7-39.7h-23.7V172.4z}
                 svg{M88.7,56.8c0,5.5-4.5,10.1-10.1,10.1c-5.6,0-10.1-4.6-10.1-10.1c0-5.6,4.5-10.1,10.1-10.1C84.2,46.7,88.7,51.3,88.7,56.8z};}}
\newcommand\orcid[1]{\href{https://orcid.org/#1}{\mbox{\scalerel*{
\begin{tikzpicture}[yscale=-1,transform shape]
\pic{orcidlogo};
\end{tikzpicture}
}{|}}}}
\title{General parametrization of wormhole spacetimes and its application to shadows and quasinormal modes}

\author[a,b,c]{
Kirill A.~Bronnikov \orcid{0000-0001-9392-7558},}
\author[b,d]{Roman A.~Konoplya \orcid{0000-0003-1343-9584},}
\author{\\and}
\author[d]{Thomas D.~Pappas  \orcid{0000-0003-2186-357X}}

\emailAdd{kb20@yandex.ru}
\emailAdd{roman.konoplya@gmail.com}
\emailAdd{thomas.pappas@physics.slu.cz}

\affiliation[a]{Center for Gravitation and Fundamental Metrology, All-Russian Research Institute of
Metrological Service, Ozyornaya 46, Moscow 119361, Russia}

\affiliation[b]{Institute of Gravitation and Cosmology, Peoples’ Friendship University of Russia (RUDN University),
ulitsa Miklukho-Maklaya 6, Moscow 117198, Russia}

\affiliation[c]{National Research Nuclear University ``MEPhI'' (Moscow Engineering Physics Institute),
Kashirskoe shosse 31, Moscow 115409, Russia}

\affiliation[d]{Research Centre for Theoretical Physics and Astrophysics, Institute of Physics, Silesian University in Opava, Bezru\v{c}ovo n\'am\v{e}st\'i 13, CZ-746 01 Opava, Czech Republic}

\abstract{
The general parametrization for spacetimes of spherically symmetric Lorentzian, traversable wormholes in an arbitrary metric theory of gravity is presented. The parametrization is similar in spirit to the post-Newtonian parametrized formalism, but with validity that extends beyond the weak field region and covers the whole space. Our method is based on a continued-fraction expansion in terms of a compactified radial coordinate. Calculations of shadows and quasinormal modes for various examples of parametrization of known wormhole metrics that we have performed show that, for most cases, the parametrization provides excellent accuracy already at the first order. Therefore, only a few parameters are dominant and important for finding potentially observable quantities in a wormhole background. We have also extended the analysis to the regime of slow rotation.
}

\begin{document}
\maketitle
\vspace{0.8cm}

\vspace{0.5cm}
%===============================================
\section{Introduction \label{sec:intro}}
%===============================================

Wormholes (WHs) belong to a special class of solutions to the Einstein equations representing tunnel-like structures which connect spatially separated regions or even different universes. The first description of such a geometry appeared as early as in 1916 in the paper by Flamm \cite{Flamm:1916} in a study of the spatial part of the Schwarzschild metric. A WH geometry also emerged during the effort of Einstein and Rosen in the mid-1930s towards a geometric description of elementary particles \cite{Einstein:1935tc}. However, the foundations for our current understanding of WHs have been laid in the seminal work by Morris and Thorne (MT) in the late 1980s \cite{Morris:1988cz} where they investigated the conditions for traversability of these speculative objects by human travelers.

As MT's work and subsequent study has revealed, WHs come with a number of problems, such as the necessity of exotic matter in order to keep the WH throat open \cite{Morris:1988cz}, or dynamical instability \cite{Bronnikov:2002sf,Gonzalez:2008wd,Bronnikov:2011if,Bronnikov:2012ch,Bronnikov:2018vbs,Cuyubamba:2018jdl} of WHs. By now, there is no noncontradicting model of a traversable, Lorentzian WH, which is dynamically stable, does not require exotic matter for its existence and follows from some fundamental theoretical principles. Nevertheless, even a hypothetical possibility to create WHs in a distant future experiment looks attractive and justifies the effort towards a further study of various WH solutions.

Since there is no fully satisfactory WH model, an appealing question is how to describe the geometry of WHs in a context as general as possible. A general parametrization of a WH geometry made in the spirit of the parametrized post-Newtonian (PPN) formalism, but valid in the whole space from the throat to infinity, could be a solution. This would allow one to constrain possible WH geometries in the current and future experiments via constraints on appropriate parameters of the parametrization.

This kind of program has been recently fulfilled by Rezzolla and Zhidenko (RZ) \cite{Rezzolla:2014mua} who suggested a parametrization of arbitrary static, spherically symmetric black hole (BH) metrics convenient for comparison with observations,
independently of a theory of gravity. The RZ parametrization enables one to approximate any
sufficiently smooth BH metric with any prescribed accuracy using a minimum possible number of numerical parameters. This parametrization was further extended to axially symmetric BHs in \cite{Konoplya:2016jvv}, to higher-dimensional BHs in \cite{Konoplya:2020kqb} and applied to analytical representation of various numerical BH solutions in \cite{Younsi:2016azx,Kokkotas:2017zwt,Kokkotas:2017ymc,Hennigar:2018hza,Konoplya:2018arm,Konoplya:2019goy,Konoplya:2019fpy,Khodabakhshi:2020hny}.

The method \cite{Rezzolla:2014mua,Konoplya:2016jvv} is based on continued-fraction expansions of the metric functions in the radial direction in terms of a compact coordinate and simultaneous expansion in the polar direction in terms of $\cos \theta$ around the equatorial plane. The continued-fraction expansion provides the superior convergence and clear hierarchy of parameters. The latter is necessary to constrain effectively the allowed geometries of a compact object from experiments.

It is then straightforward to formulate a similar kind of parametrization for WHs. Instead of the event horizon radius used as a natural length scale for BHs, for a WH such a natural length scale is given by its throat radius, or, if there are multiple throats (as is the case in some models discussed in the literature), it makes sense to speak of the throat closest to the observers, or, in other words, the external one, outside which we can assume that the spherical radius is a growing function of some reasonably chosen radial coordinate. The radius $r_0$ of this throat can be well used as a fixed length parameter for WHs instead of the horizon radius used when discussing BH metrics.

Here we will construct a general parametrization for Lorentzian, traversable, asymptotically flat, spherically symmetric WHs, not necessarily symmetric relative to the throat, which is independent on the background metric theory of gravity. We will further extend this general parametrization to axial symmetry in the slow rotation regime. We will show that once the metric functions relatively slowly approach their asymptotic values, only a few dominant parameters of the parametrization determine the behavior of potentially observable quantities around WHs with high accuracy. In the slow rotation regime, this general form of the metric, independent of the gravitational theory, has the following form:
\ba
ds^2 &=&-f(r) dt^2 +\frac{1}{h(r)} dr^2 -\frac{4m \widetilde{\alpha}}{r} \sin^2\theta dt d\phi + r^2 \left( d\theta^2 +\sin^2\theta d \phi^2 \right)\,,\\
f(r) &=&1-\frac{r_0 \left(1+\eps \right)}{r}+ \frac{r_0^3 \left(a_1+f_0+\eps \right)}{r^3}-\frac{r_0^4\, a_1}{r^4}\,,\\
h(r) &=&1-\frac{r_0 \left(1+\eps \right)}{r}+ \frac{r_0^2 \left(b_1+h_0+\eps \right)}{r^2}-\frac{r_0^3\, b_1}{r^3}\,,
\ea
where $r_{0}$ is the location of the WH's throat, $ \widetilde{\alpha}$ is the rotation parameter, $h_{0}$, $f_{0}$ are the values of the metric functions at the location of the throat, and $\epsilon$, $a_1$, $b_1$ are parameters of deformation. Whenever more accuracy is required, two more parameters of deformation are added. This form of the WH metric could be used for testing various astrophysical phenomena, such as accretion of matter, shadows, various types of radiation phenomena, and further constraining of the allowed geometries of wormholes via constraining the appropriate parameters.

Another possible application of our parametrization is a derivation of analytical approximations for numerical WH solutions \cite{Antoniou:2019awm,Ibadov:2020btp}, and for this kind of work we suggest various approaches to the construction of alternative parametrizations which take into account a convenient choice of the coordinate system under consideration and the behavior of 
the metric near the throat. We calculate the potentially observable characteristics, such as WH shadows and quasinormal modes, and see that those observable values for the parametrized approximation of WH metrics have a relative error (as compared to exact solutions) which ranges from about a small fraction of one percent to, in the worst cases, a few percent already at the first order of the expansion in the radial direction. The second-order approximation provides always an excellent accuracy if it does not simply coincide with the exact solution.

The paper is organized as follows. In Sec.~\ref{Sec:II} we summarize the general information about WHs and various choices of coordinate systems used for their description. Section~\ref{Sec:III} is devoted to the construction of a general parametrization for spherically symmetric and axially symmetric WHs in the slow rotation regime. Section~\ref{Sec:IV} tests this parametrization using a great number of examples of WH metrics that can be recast to the MT frame. For these cases, the radial coordinate is conceptually identical to the one used in the parametrization of BH metrics. In Sec~\ref{Sec:V} we deal with the parametrization of WHs that are not in the MT frame and develop optimized parametrizations via different choices of a compact radial coordinate that take into consideration the behavior of the metric near the throat. In Sec.~\ref{Sec:VΙ} we test the accuracy of the parametrized description via
the calculation of the radii of shadows of WHs and their quasinormal spectra. Finally, in Sec.~\ref{Sec:conclusions} we summarize the obtained results and discuss open questions. 

\vspace{0.5cm}
%===============================================
\section{Wormholes in different coordinates \label{Sec:II}}
%===============================================

The general metric describing an arbitrary static, spherically symmetric geometry may 
be written in the form
\beq
ds^2=-f(r) dt^2 +\frac{1}{h(r)} dr^2 +K^2(r) d\Omega^2 \,,
\label{gen_l_e}
\eeq
where $d\Omega^2 =\left( d\theta^2+ \sin^2 \theta\, d\phi^2 \right)$ is the line element on a two-dimensional unit sphere, and $r$ is an arbitrary radial coordinate, whose specific choice can be made for convenience. Only two of the three metric functions $f(r)$, $h(r)$, $K(r)$ are independent, and upon using appropriate transformations of the radial coordinate, any metric can be cast in the form where the circumferential radius $K(r)$ satisfies $K(r)=r$, albeit this might not always be feasible analytically. In general, the area of the 
sphere at radial coordinate $r$ is $A(r)=4\pi\, K^2(r)$.

It is said that the metric \rf{gen_l_e} describes a (traversable, Lorentzian) WH if the following two conditions are met. First, the circumferential radius has at least one minimum 
$K_0 \equiv K(r_0)$ at some value of the radial coordinate $r_0$, and $K(r)$ should be large as compared to $K_0$ on both sides from this minimum. Then, $K_0$ corresponds to the radius of the WH throat, and $r_0$ is its location. At the throat, the area of  the constant $r$ sphere is 
minimized, and this allows one to determine $r_0$ via the condition
\beq
A'(r_0)=0 \rightarrow K'(r_0)=0\,.
\eeq
Second, the functions $f(r)$ and $h(r)$ are regular and positive in a range of $r$ containing the throat and values of $r$ on both sides from the throat such that $K(r) \gg K(r_0)$. Such a definition includes both asymptotically flat or AdS WHs and those containing horizons far from the throat, for example, asymptotically de Sitter ones. In this work we will consider asymptotically flat WHs such that, as $r$ tends to some $r = r_\infty$,
\beq             
f \rightarrow 1\,, \qquad h \left(\frac{dK}{dr} \right)^2 \rightarrow 1\,.
\label{asflat}
\eeq  
The WH metrics discussed in the literature are written using different choices of the coordinate $r$, and we here enumerate three  of them:
\begin{enumerate}[(i)]
\item
The curvature coordinate defined by $r = K$, such that the coordinate is identified with the spherical radius. In this coordinate system, the second condition \rf{asflat} reads $h(r) \to 1$ as $r \to \infty$. For WHs this choice of coordinate is unnatural	because a minimum of $K$, i.e., a throat,	is a coordinate singularity, $h^{-1} \to \infty$. However, it is often used since it is rather intuitively clear and simplifies the gravitational field equations in the presence of some (but not all) material sources of gravity.
\item
	The ``quasiglobal'' coordinate $r$ such that $f(r)/ h(r) \equiv 1$. This coordinate is 		
	especially convenient for describing BH horizons but is also used in many 
     WH solutions. We can recall that the most well-known solutions of general relativity (\Scz,
	\RN, (A)dS) are most often written in terms of $r$ which is
	simultaneously a curvature and quasiglobal coordinate. Since in \rf{asflat} the 
	first condition requires $f \to 1$, the second one reduces to $|dK/dr| \to 1$.
\item
	The Gaussian, or proper length coordinate $r = l$, which is defined in such a way that 
	$h \equiv 1$. The second condition \rf{asflat} reads: $|dK/dl| \to 1$ as $l \to \infty$.
\end{enumerate}

\noindent In terms of the curvature coordinate, a very common frame where many 
WH metrics are written is the one introduced by Morris and Thorne \cite{Morris:1988cz} 
\beq
ds^2=-e^{2\Phi(r)}dt^2+\left(1-\frac{b(r)}{r} \right)^{-1} dr^2 +r^2 d\Omega^2\,.
\label{MT_ansatz_le}
\eeq
The first metric function $\Phi(r)$ is the so-called ``redshift'' function that determines 
the redshift and tidal forces in the WH spacetime. The absence of event horizons demands 
that $\Phi(r)$ should be finite everywhere. The second function $b(r)$ is called the ``shape'' function since it indirectly determines the spatial shape of the WH in its embedding diagram representation. It should satisfy the so-called flair-out conditions on the throat, i.e., $b(r_0)=r_0$ and $b'(r_0)<1$ while $b(r)<r$ for $r \neq r_0$, which are actually a reformulation of the condition that $K(r)$ has a minimum in the more general representation \rf{gen_l_e}.
The aforementioned conditions on $h(r)$ imply that the location of the throat $r_0$ in the MT frame is given as a root of the equation
\beq
h(r) = \left(1-\frac{b(r)}{r} \right)=0\,,  \label{MT_r0}
\eeq
while $e^{\Phi(r_0)} > 0$, and the curvature coordinate is defined for $r \in [r_0,\infty)$. 
Finally, it should be noted that asymptotic flatness demands that 
$b(r)/r \rightarrow 0$ as $r \rightarrow \infty$, which translates to $h(r) \rightarrow 1$.

\vspace{0.5cm}
%===============================================
\section{Continued-fraction parametrizations of asymptotically flat metrics \label{Sec:III}}
%===============================================

In this section, we will briefly review the parametrization of spherically symmetric BHs suggested in \cite{Rezzolla:2014mua}, and then we will see which modifications of this approach are required when going over to WH geometries.

% ----------------------------------------------------------------------
\subsection{Overview of the Rezzolla-Zhidenko method}
% ----------------------------------------------------------------------

\noindent 
The Rezzolla-Zhidenko parametrization is build around a dimensionless compact coordinate (DCC) defined by
\beq
x(r)\equiv 1-\frac{r_0}{r}\,,			\label{RZ_DCC}
\eeq
where $r_0$ is the location of the outer event horizon of the BH determined via the condition $f(r_0)=0$. If $K^2(r)=r^2$, i.e., one works with the curvature coordinate, then $r_0$ is also the radius of the outer event horizon.
In terms of this DCC, the following parametrization equations are introduced:
\ba
f(r)&=&\widetilde{A}(x)\,,\label{RZ_param_eq_A}
\\
\frac{1}{h(r)}&=&\frac{\widetilde{B}(x)}{\widetilde{A}(x)}\,,
\label{RZ_param_eq_B}
\ea
where the parametrization functions $\widetilde{A}(x)$ and $\widetilde{B}(x)$ are defined as
\ba
\widetilde{A}(x) &\equiv& x \left[ 1-\epsilon (1-x)+(a_0-\epsilon)(1-x)^2
+\frac{a_1 }{1+\frac{a_2 x}{1+\frac{a_3 x}{\ldots}}}(1-x)^3  \right]\,,      \label{RZ_param_fun_2}
\\
\widetilde{B}(x) &\equiv&\left[ 1+b_0 (1-x) 
+ \frac{b_1 }{1+\frac{b_2 x}{1+\frac{b_3 x}{\ldots}}}(1-x)^2 \right] ^2\,.   \label{RZ_param_fun_3}
\ea
There are three asymptotic parameters in total, namely, $(\eps, a_0,b_0)$, which are determined via the expansions of the parametrization equations at spatial infinity ($x=1$). The remaining parameters $(a_1,a_2,\ldots ,b_1,b_2,\ldots)$ are the ``near-field'' parameters and are determined by the corresponding expansions at the location of the event horizon ($x=0$).

The observational constraints on the asymptotic parameters $(\eps, a_0,b_0)$ are imposed via the PPN expansions \cite{Will:2005va,Will:2014kxa}
\ba
f(r)&=&1-\frac{2M}{r}+(\beta -\gamma) \frac{2 M^2}{r^2}+\mathcal{O}\left( \frac{1}{r^3} \right)\, \nonumber\label{PPN_exp_gtt}
\\
&=& 1- \frac{2M}{r_0} \left(1-x \right) + \left( \beta-\gamma \right) \frac{2 M^2}{r_0^2} \left( 1-x \right)^2 +\mathcal{O}\left( \left(1-x \right)^3 \right)\,,
\ea
and
\beq
\frac{1}{h(r)}=1+\gamma \frac{2 M}{r}+\mathcal{O}\left( \frac{1}{r^2} \right)
= 1+\gamma \frac{2M}{r_0} (1-x)+\mathcal{O}\left( \left(1-x \right)^2 \right)\,.             	
\label{PPN_exp_grr}
\eeq
Notice that the highest-order PPN constraints on the metric are of the order 
$\mathcal{O}\left( \left(1-x \right)^2 \right)$ for the expansion of the $g_{tt}$ metric 
component and of $\mathcal{O}\left( \left(1-x \right) \right)$ for $g_{rr}$. Consequently, this means that we can impose independent constraints for up to three asymptotic parameters in the parametrization of $f(r)$ and up to two for $h(r)$. The values of the PPN parameters 
$\beta$ and $\gamma$ are observationally constrained to be \cite{Will:2005va,Will:2014kxa}
\beq
\left| \beta -1 \right| \lesssim 2.3 \times 10^{-4}\,,\qquad 
\left| \gamma -1 \right| \lesssim 2.3 \times 10^{-5}\,.         \label{gamma_beta_constr}
\eeq
The expansion of the parametrization functions~\eqref{RZ_param_fun_2} and~\eqref{RZ_param_fun_3} at $x=1$ are respectively
\beq
\widetilde{A}(x) =1-(1+\eps) (1-x)+a_0 (1-x)^2+\mathcal{O}\left( \left(1-x \right)^3 \right)\,,
\label{A(x)_inf_exp}
\eeq
and
\beq
\frac{\widetilde{B}(x)}{\widetilde{A}(x)} =1+(1+2 b_0+\eps) (1-x)+\mathcal{O}\left( \left(1-x \right)^2 \right)\,.
\label{B(x)_inf_exp}
\eeq
Then, the comparison of the expansions~\eqref{PPN_exp_gtt},~\eqref{A(x)_inf_exp} and~\eqref{PPN_exp_grr},~\eqref{B(x)_inf_exp} imposes the observational constraints
\beq
\eps =\frac{2 M}{r_0}-1\,,\quad a_0=\frac{2M^2}{r_0^2} \left( \beta-\gamma \right)\,,
\eeq
and
\beq
(1+2 b_0+\eps)= \gamma \frac{2M}{r_0}\,\quad \Rightarrow \quad b_0= \frac{M}{r_0} \left(\gamma-1 \right)\,.
\eeq
One then concludes that viable BH solutions must have $a_0 \simeq 0$ and $b_0 \simeq 0$ in order to comply with the observations according to \eqref{gamma_beta_constr}.

% -------------------------------------------------------------------
\subsection{Static, spherically symmetric wormholes}
% -------------------------------------------------------------------

\noindent We are now going to extend the above prescription to accommodate the parametrization of WH metrics. This can be achieved by appropriately modifying the parametrization equations~\eqref{RZ_param_eq_A},~\eqref{RZ_param_eq_B} and the parametrization functions~\eqref{RZ_param_fun_2},~\eqref{RZ_param_fun_3}. As we have discussed in Sec.~\ref{Sec:II}, the location of the WH throat $r_0$ is defined by the condition $K'(r_0)=0$ or in the MT frame by $h(r_0)=0$. We are then interested in a parametrization of the WH metric in the region $r \in [r_0,\infty)$. As in the case of BHs, it is sufficient to parametrize only two metric functions under an appropriate choice of the radial coordinate.

Following the RZ approach, we define the DCC as in Eq.~\eqref{RZ_DCC}, thus the DCC
maps the interval $r \in [r_0,\infty)$ to the compact range $x \in [0,1]$. The next step is to introduce two extra near-field parameters $f_0, h_0$ to account for the fact that the WH metric functions
$f(r)$ and $h(r)$ in principle attain nonvanishing values at $r=r_0$. This is to be contrasted to
the BH case where $f(r_0)=h(r_0)=0$ at the location of the outer event horizon, and so the parametrization function $\widetilde{A}(x)$~\eqref{RZ_param_fun_2} must also vanish at 
$r_0$ or equivalently at $x=0$. In terms of the DCC~\eqref{RZ_DCC} we define the following parametrization equations:
\ba
f(r)&=& A(x)\,,\label{WH_param_eq_A}\\
h(r)&=& B(x)\,,
\label{WH_param_eq_B}
\ea
with the parametrization functions given by
\ba
A(x) &\equiv& f_0+x \left[ \left(1-f_0 \right)-\left(\epsilon +f_0 \right) (1-x)+(a_0-\epsilon-f_0)(1-x)^2+\frac{a_1 (1-x)^3}{1+\frac{a_2 x}{1+\frac{a_3 x}{\ldots}}} \right]\,,\label{WH_param_fun_A}\\
B(x) &\equiv& h_0+x\left[ \left(1 -h_0\right)-\left(b_0+h_0 \right) (1-x) + \frac{b_1 (1-x)^2 }{1+\frac{b_2 x}{1+\frac{b_3 x}{\ldots}}} \right] \,.
\label{WH_param_fun_B}
\ea
To impose observational constraints on the asymptotic parameters, we consider 
the expansions at $x=1$
\ba
A(x) &=& 1-\left(1+\eps \right) \left(1-x \right)+a_0 \left(1-x \right)^2+\mathcal{O}\left( \left(1-x\right)^3 \right)\,,\label{A(x)_WH_inf_exp}\\
\frac{1}{B(x)} &=& 1+\left(1+b_0 \right)\left( 1-x\right)+\mathcal{O}\left( \left(1-x \right)^2 \right)\,.\label{B(x)_WH_inf_exp}
\ea
A comparison with the PPN expansions~\eqref{PPN_exp_gtt} and~\eqref{PPN_exp_grr} 
yields the constraints
\beq
\eps =\frac{2 M}{r_0}-1\,,\quad a_0=\frac{2M^2}{r_0^2} \left( \beta-\gamma \right)\,,
\eeq
and
\beq
b_0= \gamma \frac{2M}{r_0}-1\,\quad \Rightarrow \quad b_0= \gamma \left( \eps+1 \right)-1\,.    \label{PPN_b0}
\eeq
Thus, in our parametrization, observationally viable solutions that comply with the PPN constraints~\eqref{gamma_beta_constr}, must be characterized by $a_0 \simeq 0$ 
and $b_0 \simeq \eps$.

% ---------------------------------------------------------------------
\subsection{Wormholes in the slow rotation approximation}
% ---------------------------------------------------------------------

\noindent The stationary, axisymmetric generalization of the MT wormhole \cite{Morris:1988cz} was found in \cite{Teo:1998dp}, and the line element is given by
\beq
ds^2=-fdt^2+\frac{1}{h}dr^2+ K^2 \left[ d \theta^2+\sin^2 \theta \left( d \phi -\omega dt \right)^2\right]\,,
\label{Teo_WH}
\eeq
where the metric functions $f,h,K$ and $\omega$ depend only on $r$ and $\theta$ and are regular on the symmetry axis $\theta=0,\pi$. In terms of the rotation parameter 
$\widetilde{\alpha} \equiv J/m$, where $J$ is the angular momentum of the WH as measured
by an asymptotic observer and $m$ is its mass, the slow rotation approximation is obtained by expanding the metric functions in terms of the dimensionless parameter $\widetilde{\alpha}/m \ll 1$. Up to the linear order in $\widetilde{\alpha}$ we have
\beq
ds^2 =-f(r) dt^2 +\frac{1}{h(r)} dr^2 + K^2(r) d \Omega^2 +g(r)  \sin^2 \theta dt d\phi +\mathcal{O}(\widetilde{\alpha}^2)\,,
\label{SR_le}
\eeq
where it is assumed that all functions depend only on the radial coordinate,
 and we have defined the metric function
\beq
g(r) \equiv -2 \omega(r) K^2(r)\,,
\eeq
that involves the angular velocity metric function $\omega(r)$. The latter depends on the rotation parameter linearly and should exhibit asymptotically the following fall-off behavior:
\beq
\omega(r)=\frac{2 J}{r^3}+\mathcal{O}\left(\frac{1}{r^4} \right)\,.
\eeq
Furthermore, asymptotic flatness requires that $K^2(r) \rightarrow r^2$ as $r \rightarrow \infty$  and thus $g(r)$ has the asymptotic behavior
\ba
 g(r)&=&-\frac{4 J}{r}+\mathcal{O}\left(\frac{1}{r^2} \right)\,,\label{g(r)_asym_exp_r}\\
 &=&-\frac{4 J}{r_0} \left(1-x \right) +\mathcal{O}\left( \left(1-x \right)^2 \right)\,.
 \label{g(r)_asym_exp}
\ea
By analogy with the previous sections, in terms of the DCC of Eq.~\eqref{RZ_DCC}, a parametrization function can be introduced for $g(r)$ with a single asymptotic parameter $c_0$ and a tower of near-field parameters $(g_0,c_1,c_2,\ldots)$ as
\beq
C(x)\equiv g_{0} +x \left[ -g_0- \left(c_0+g_0 \right)\left(1-x \right)+\frac{c_1 \left(1-x \right)^2}{1+\frac{c_2 x}{1+\frac{c_3 x}{1+\ldots}}} \right]\,.
\label{WH_param_fun_C}
\eeq
The parametrization equation is defined as 
\beq
g(r) \equiv C(x) \,,
\label{WH_param_eq_C}
\eeq
while the asymptotic expansion of Eq.~\eqref{WH_param_fun_C} at infinity reads
\beq
C(x)=- c_0 \left (1-x \right)+ \mathcal{O}\left( \left( 1-x\right)^2 \right)\,.
\label{C(x)_WH_inf_exp}
\eeq
Consequently, a comparison with Eq.~\eqref{g(r)_asym_exp} reveals that the asymptotic parameter $c_0$ is associated with the angular momentum via $c_0= 4J/r_0$.

\vspace{0.5cm}
%=======================================================
\section{Parametrization of Wormholes in the Morris-Thorne frame \label{Sec:IV}}
%=======================================================

\noindent In this section, we are going to test our method by parametrizing different types of analytical WH metrics that can be brought to the MT frame~\eqref{MT_ansatz_le} and consequently are written in terms of the curvature coordinate as defined in Sec.~\ref{Sec:II}. Once a parametrization is obtained, we compute the error of the parametrized metrics at different orders in the continued-fraction expansion  and demonstrate the high accuracy of the approximation already at the lowest orders as well as its quick convergence.

% ------------------------------------------------------------------------------------
\subsection{The Bronnikov-Kim II braneworld wormhole solution}

\begin{figure*}[ht!]
%\centering
\includegraphics[width=0.5\linewidth]{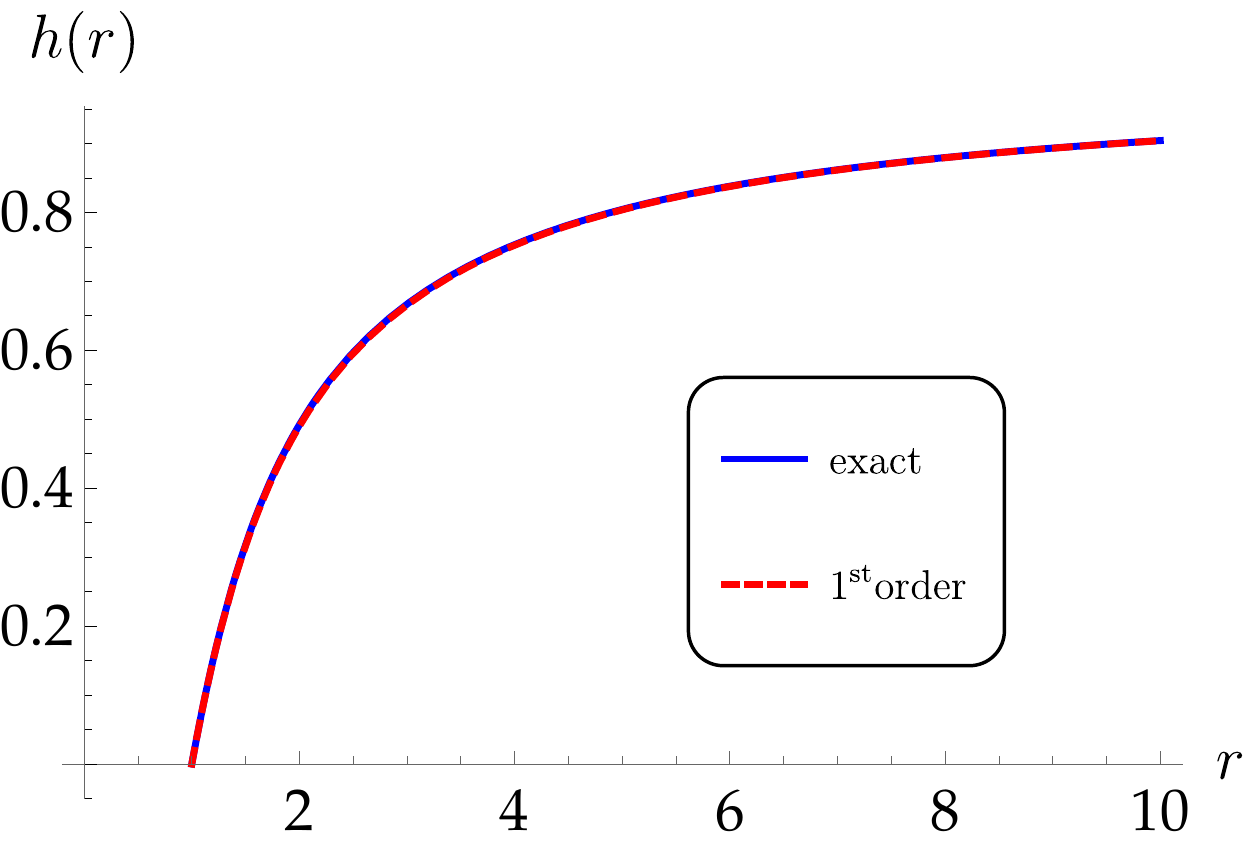}
\includegraphics[width=0.5\linewidth]{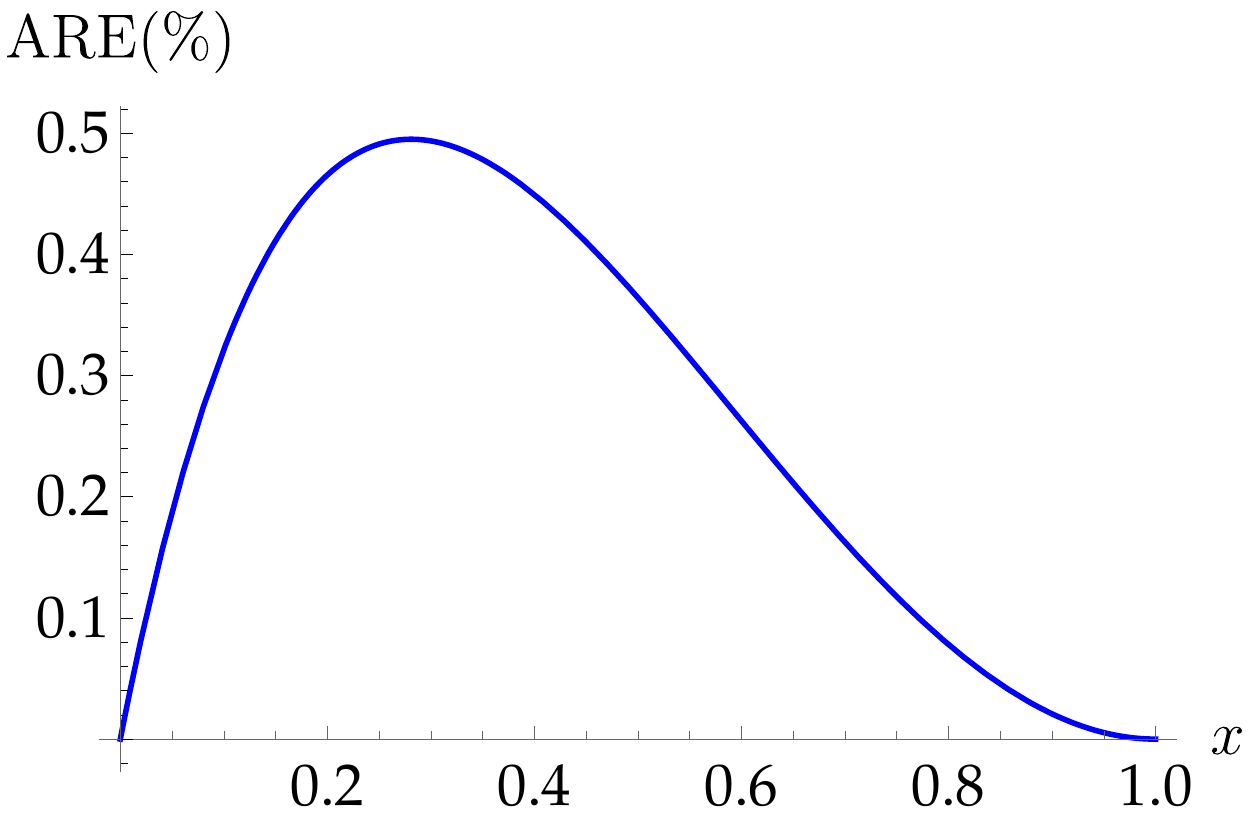}
\caption{For $p \equiv a/r_0 =0.5$. Left panel: the exact metric function  $h(r)$~\eqref{BKzeromassg11} (blue curve) and its first-order approximation (red dashed curve). 
Right panel: the percentage of the absolute relative error in terms of the dimensionless compact coordinate $x \in [0,1]$.
}\label{fig:BK2h_r_ARE}
\end{figure*}

\noindent As the first example, we consider a braneworld solution 
\cite{Bronnikov:2002rn,Bronnikov:2003gx,Bronnikov:2019sbx}
that exhibits BH and WH branches with zero
Schwarzschild mass and is described by the following line element:
\beq
ds^2=-\left(1-\frac{a^2}{r^2} \right)dt^2+\left(1-\frac{a^2}{r^2} \right)^{-1} \left(1+\frac{C-a}{\sqrt{2 r^2-a^2}} \right)^{-1} dr^2+r^2 d\Omega^2\,. \label{BKzeromass}
\eeq
The WH branch of the metric should be free of event horizons and this requirement translates to $f(r) \equiv - g_{tt}(r)>0\, \forall \, r \in [r_0,\infty)$, where $r_0>0$ is the location of the WH throat. Consequently, the parameter $a$ is constrained by
\beq
f(r_0)=\left(1-\frac{a^2}{r_0^2} \right) \geqslant 0  \Rightarrow r_0 \geqslant a\,,
\eeq
with the equality $a=r_0$ corresponding to the WH/BH threshold where $r_0$ is then identified with the location of the (double) BH event horizon. Since the metric~\eqref{BKzeromass} is of the MT type, the location of the WH throat is determined by the condition $h(r_0)=0$~\eqref{MT_r0} via which we find that the parameter $C$ should satisfy 
\beq
C =a-\sqrt{2 r_0^2-a^2} \leqslant 0 \,,
\eeq
where in the WH branch $C<0$ and at the WH/BH threshold $C=0$. Upon substituting the last equation into Eq.~\eqref{BKzeromass} we may express the metric function $h(r)$ entirely in terms of $r_0$ and $a$ as
\beq
h(r)=\left(1-\frac{a^2}{r^2} \right) \left( 1-\frac{\sqrt{2 r_0^2-a^2}}{\sqrt{2 r^2-a^2}}\right)\,.
\label{BKzeromassg11}
\eeq
We begin with the parametrization of the $f(r)$ metric function in terms of the dimensionless compact coordinate $x(r)$ given by Eq.~\eqref{RZ_DCC}. As anticipated by the polynomial 
form of the metric function, the comparison of the expansions of the parametrization equation~\eqref{WH_param_eq_A} at the boundaries $x=1$ and $x=0$ yields an exact parametrization with the values of the expansion parameters (EPs) being
\beq
\eps=-1\,,\quad a_0=-\frac{a^2}{r_0^2} \,,\quad f_0=1-\frac{a^2}{r_0^2}\,,\quad a_i =0\quad \forall i \geqslant 1\,.
\eeq
The parametrization of $h(r)$ is not exact, and the corresponding expansions of~\eqref{WH_param_eq_B} at the boundaries yield
\ba
b_0&=&\sqrt{1-\frac{a^2}{2 r_0^2}}-1 \,,\quad h_0=0\,,\quad b_1 =b_0+ \frac{a^2}{a^2+2r_0^2}\,,\\
b_2 &=& \frac{2 a^4 \left(1+b_0 \right)-a^2 \left(7+8 b_0 \right) r_0^2+8 b_0 r_0^4}{b_1 \left( a^2 -2\, r_0^2 \right)^2}-3\,,
\ea
where we have only listed the first few EPs which have compact forms. To obtain the 
approximation of the metric to the $i$th order, the continued-fraction expansions in the parametrization functions~\eqref{WH_param_fun_A}-\eqref{WH_param_fun_B} are to be
truncated at the $i$th order by setting the $(i+1)$th expansion parameter equal to zero. As an example, in the left panel of Fig.~\eqref{fig:BK2h_r_ARE} we plot the exact metric function 
$h(r)$ as given in Eq.~\eqref{BKzeromassg11} and its approximation $h_{\rm app}(r)$ 
at first order for $p \equiv a/r_0= 0.5$, where we have set $b_2 =0$. In the right panel 
of the same figure, and for the same value of the parameter $p$, we plot, in terms of the 
compact coordinate~\eqref{RZ_DCC}, the percentage of the absolute relative error 
(ARE) given by
\beq
\text{ARE} \equiv 100\left|\frac{h(r)-h_{\rm app}(r)}{h(r)} \right| \% \,.
\eeq
The radial profile of the ARE is the typical profile obtained in all cases, with the error
vanishing at the boundaries of the parametrization range $r \in [r_0,\infty) \rightarrow x \in [0,1]$ while exhibiting a global maximum at some intermediate value of $x(r)$ that emerges at a value of $r$ not far from the location of the throat.

To test the convergence of the parametrization, in Table~\eqref{table:BKzeromass} we give the maximum percentage value of the ARE for the approximation $h_{\rm app}(r)$ of $h(r)$ at various orders in the continued-fraction expansion and for different values of the dimensionless
parameter $p \equiv a/r_0 \leqslant 1$ with the WH/BH threshold corresponding to $p=1$.

\begin{table}[H]
\center{
\caption{The maximum absolute relative error in percents between the exact metric $h(r)$~\eqref{BKzeromassg11} and its approximation at various orders  in terms of the dimensionless parameter $p \equiv a/r_0 \leqslant 1$. The WH/BH threshold corresponds to $p=1$.}
\begin{tabular}{|c|c|c|c|c|c|c|}
 \hline
 order
  &  $p=0.1$ & $p=0.3$ & $p=0.5$ & $p=0.7$ & $p=0.8$ & $p=0.99$ \\
 \hline\hline
  1 & 0.00063 & 0.05460 & 0.49509 & 2.53408 & 5.39226 & 31.12707  \\
 \hline
  2 & 0.00010 & 0.00840 & 0.07270 & 0.34042 & 0.67203 & 2.49612 \\
 \hline
  3 & 0.00001 & 0.00118 & 0.00829 & 0.02370 & 0.02622 & 0.13332  \\
 \hline
   4 &  $\mathcal{O}(10^{-8})$ & 0.00004 & 0.00093 &  0.00883 &  0.02312 &  0.12947  \\
 \hline
\end{tabular}
\label{table:BKzeromass}
}
\end{table}

As the elements of Table~\eqref{table:BKzeromass} verify, when  $p \equiv a/r_0$ is smaller than about $0.5$, our parametrization provides a very accurate approximation of the metric even at the first order, with a maximum absolute relative error (MARE) below $1 \%$. For larger values of $p$ and as the WH/BH threshold is approached, only the first-order MARE becomes large with the higher-order approximations preserving a negligible MARE with values well below $1 \%$. Furthermore, we notice that in the limit $p \rightarrow 1$, the approximation remains convergent albeit at a slower rate.

% ----------------------------------------------------------------
\subsection{The Casadio-Fabbri-Mazzacurati metric}
% ----------------------------------------------------------------

\noindent In the context of the braneworld scenario, the search for BH and WH solutions led to the following very interesting geometry \cite{Germani:2001du,Casadio:2001jg,Bronnikov:2002rn,Bronnikov:2003gx}:
\beq
ds^2=-\left(1-\frac{2m}{r} \right)dt^2+\frac{1-\frac{3m}{2r}}{\left(1-\frac{2m}{r} \right) \left(1-\frac{r_0}{r} \right)}dr^2+r^2 d\Omega^2\,,
\label{BKCFM_r}
\eeq
that describes different objects depending on the values of $r_0>0$ and $m$ (we take $m>0$). It is convenient to introduce the dimensionless parameter $p \equiv r_0/m >0$ in terms of which we have the following branches \cite{Bronnikov:2002rn}:
\vspace{0.5 cm}
\begin{enumerate}
    \item{$p \in (2,\infty]$ : Symmetric traversable WH.}
    \item{$p=2$ : WH/BH threshold.}
    \item{$p \in (\frac{3}{2},2)$ : Regular BH.}
    \item{$p=\frac{3}{2}$ : Schwarzschild BH.}
    \item{$p \in (0,\frac{3}{2})$ : A Schwarzschild-like BH structure with a spacelike 
    curvature singularity located at $r=\frac{3}{2}m$.}
\end{enumerate}
\vspace{0.5 cm}
The WH branch of Eq.~\eqref{BKCFM_r} is of the MT type~\eqref{MT_ansatz_le} with the shape function $b(r)$ specified by the identification
\beq
h(r) = 1-\frac{b(r)}{r}=\frac{\left(1-\frac{2m}{r} \right) \left(1-\frac{r_0}{r} \right)}{1-\frac{3m}{2r}}\,.
\eeq
The location of the WH throat $r_{\rm th}$ is then determined by the condition $b(r_{\rm th})=r_{\rm th}$ and so from the last equation we find that $r_{\rm th}=r_0$. Consequently, the parametrization region in this coordinate system is $r \in [r_0,\infty)$. In terms of Eqs.~\eqref{RZ_DCC}-\eqref{WH_param_eq_B} we obtain the parametrization for $f(r)$ and $h(r)$. For the former it is exact and the values of the expansion parameters are:
\beq
\eps=\frac{2m}{r_0}-1\,,\quad a_0=0 \,,\quad f_0=1-\frac{2 m}{r_0}\,,\quad a_i =0\quad \forall i \geqslant 1\,.
\eeq
The parametrization for $h(r)$ becomes exact at the second order and the expansion parameters have the values
\beq
b_0=\frac{m}{2r_0} \,,\quad h_0=0\,,\quad b_1 =\frac{3 m^2}{6 m r_0-4r_0^2}\,,\quad b_2 =\frac{3 m}{2 r_0-3m}\,,\quad b_i =0\quad \forall i \geqslant 3\,.
\eeq
In order to test the accuracy in the first-order approximation of the metric function $h(r)$, in Table~\eqref{table:CFM_1_accu} we give the percentage of the MARE for various values of the dimensionless parameter $p \equiv r_0/m$.

\begin{table}[H]
\center{
\caption{The percentage of maximum absolute relative error between the exact metric function $h(r)$ of Eq.~\eqref{BKCFM_r} and its first-order approximation for various values of the dimensionless parameter $p \equiv r_0/m \geqslant 2$. The WH/BH threshold corresponds to $p=2$. The parametrization becomes exact at the second order.}
\begin{tabular}{|c|c|c|c|c|c|c|}
 \hline
 order
  &  $p=2.01$ & $p=2.1$ & $p=2.5$ & $p=3.5$ & $p=6$ & $p=10$ \\
 \hline\hline
  1 & 45.0447 & 24.41415 & 6.41108 & 1.13354 & 0.13316 & 0.0226685  \\
 \hline
\end{tabular}
\label{table:CFM_1_accu}
}
\end{table}

For the remainder of this section, we are going to present WH metrics that can be parametrized exactly in terms of~\eqref{RZ_DCC}-\eqref{WH_param_eq_B} with a minimal number of parameters. That is, all parameters that appear in the continued-fraction expansions 
$a_i,b_i,\, \forall i \geqslant 1$ are zero, and only the asymptotic parameters $\eps, a_0, b_0$ 
and, in some cases, $f_0,h_0$ suffice for the parametrization of the metric functions 
$f(r)$ and $h(r)$.

% --------------------------------------------------------
\subsection{Damour-Solodukhin wormhole}
% --------------------------------------------------------

\noindent A Schwarzschild-like WH metric has been proposed in \cite{Damour:2007ap} 
by deforming the Schwarzschild solution in terms of an exponentially small parameter 
$\lambda \sim e^{-4 \pi m^2}$. The line element reads
\beq
ds^2= - \left( 1- \frac{2m}{r}+\lambda^2 \right) dt^2+\left( 1- \frac{2m}{r} \right)^{-1}dr^2+r^2 d\Omega^2\,,
\label{DS_metric_static}
\eeq
and the WH throat is located at $r_0=2m$. For $\lambda \neq 0$ the metric function $f(r) \equiv -g_{tt}(r)$ is everywhere positive and no event horizon exists unless $\lambda = 0$ and the Schwarzschild black hole is recovered. In order to have a time variable that corresponds 
to the time of an asymptotic observer, we may perform the redefinition $t \rightarrow \tau$
according to $dt=\left(1+\lambda^2\right)^{-1}d \tau$, and then we have
\beq
f(r)=1-\frac{2m}{r(1+\lambda^2)}\,.
\eeq
For this geometry, the EPs which correspond to an exact parametrization via Eqs.~\eqref{RZ_DCC}-\eqref{WH_param_eq_B} are the following:
\beq
\eps=\frac{1}{1+\lambda^2}-1\,,\quad a_0= 0 \,,\quad f_0=\frac{\lambda^2}{1+\lambda^2}\,,\quad a_i =0\quad \forall i \geqslant 1\,,
\eeq
and
\beq
b_0=-\frac{1}{2} \,,\quad h_0=\frac{1}{2}\,,\quad b_i =0\quad \forall i \geqslant 1\,.
\eeq
The generalization of the metric~\eqref{DS_metric_static} to rotation has been considered in \cite{Bueno:2017hyj} by appropriately modifying the Kerr solution. At the linear order in the slow rotation approximation the line element becomes
\beq
d^2s=-\left(1-\frac{2m}{r} \right)dt^2+\left(1-\frac{2m\left(1+\lambda^2 \right)}{r} \right)^{-1}dr^2+r^2 d\Omega^2-\frac{ 4m \widetilde{\alp}}{r} \sin^2\theta dtd\phi\,.
\eeq
The location of the throat in this case is determined by the roots of the function
\beq
h(r) \equiv 1-\frac{2m\left(1+\lambda^2 \right)}{r} =0\quad \Rightarrow \quad r_0=2m\left(1+\lambda^2 \right)\,.
\eeq
So, the DCC~\eqref{RZ_DCC}, with $r_0$ defined as in the above equation, yields an exact parametrization for $h(r)$ in terms of~\eqref{WH_param_fun_B} and~\eqref{WH_param_eq_B} with all EPs equal to zero.

% ------------------------------------------------------
\subsection{Bronnikov-Kim I solution}
% ------------------------------------------------------

An interesting  WH metric was obtained as an exact solution of the 
Shiromizu-Maeda-Sasaki equations \cite{Shiromizu:1999wj} in the context of 
braneworld gravity \cite{Bronnikov:2002rn,Bronnikov:2019sbx}:
\beq
ds^2=-\left(1-\frac{2m}{r} \right)^2 dt^2 
+ \left( 1- \frac{r_0}{r} \right)^{-1} \left(1-\frac{r_1}{r} \right)^{-1} dr^2+r^2 d\Omega^2 
\quad \text{with} \quad r_1 \equiv \frac{m r_0}{r_0-m}\,.
\eeq
The location of the WH throat is at $r_0$ if $r_0 >2m$ and at $r_1$ if $r_1 >2m$. We consider the case where the throat is located at $r_0$, so the parametric range of interest to us is 
characterized by $p \equiv r_0/(2m) >1$ with a WH/BH threshold at $p=1$. Exact 
parametrizations in terms of Eqs.~\eqref{RZ_DCC}-\eqref{WH_param_eq_B} have 
the following EPs:
\beq
\eps=\frac{4m}{r_0}-1\,,\quad a_0=\frac{4m^2}{r_0^2} \,,\quad 
f_0=\frac{\left(r_0-2m \right)^2}{r_0^2}\,,\quad a_i =0\quad \forall i \geqslant 1\,,
\eeq
and
\beq
b_0=\frac{m}{r_0-m} \,,\quad h_0=0\,,\quad b_i =0\quad \forall i \geqslant 1\,.
\eeq

The solution which is identical to the above Bronnikov-Kim solution 
\cite{Bronnikov:2002rn,Bronnikov:2019sbx} up to a redefinition of the parameters was also found 
in \cite{Blazquez-Salcedo:2020czn} as an exact solution of the Einstein-Dirac-Maxwell theory.

% ---------------------------------------------------------------------
\subsection{Simpson-Visser wormhole\label{Section:SV}}
% ---------------------------------------------------------------------

\begin{figure*}[ht!]
\centering
\includegraphics[width=0.5\linewidth]{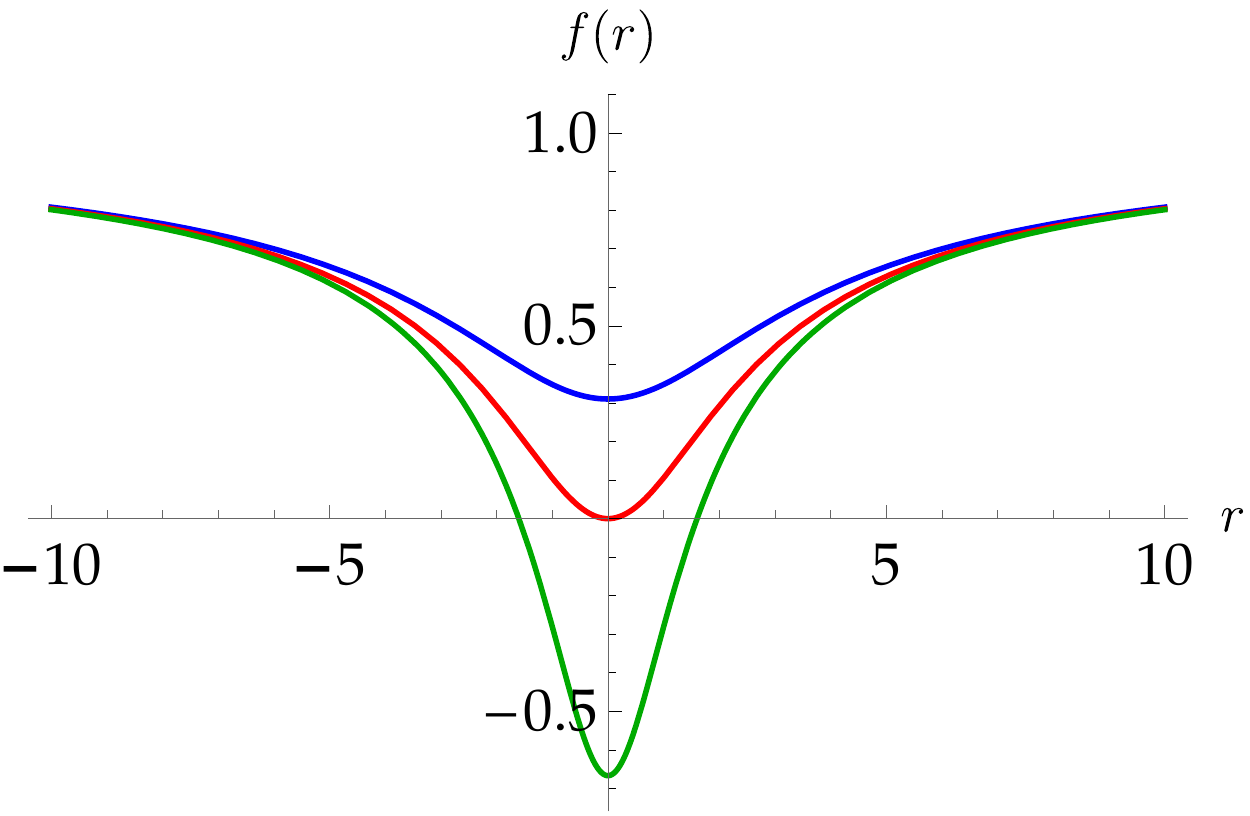}
\caption{The radial profile of the metric function $f(r) \equiv - g_{tt}(r)$ in terms of the 
quasiglobal coordinate $r \in (-\infty,\infty)$ in the three branches of the Simpson-Visser metric~\eqref{SV_met}. At $p \equiv a/m=2.9$ (blue, top), the geometry is that of a (two-way) traversable WH. At $p=2$ (red, middle) there is a horizon at the coordinate location 
$r=0$, and the geometry corresponds to an extremal regular BH, which may also be called a one-way WH with an extremal null throat. At $p=1.2$ (green, bottom), a regular BH geometry with two horizons is obtained. In the limit $p \rightarrow 0$, the Schwarzschild BH metric is recovered.}
\label{fig:SV_branches}
\end{figure*}

An interesting metric was suggested by Simpson and Visser (SV) \cite{Simpson:2018tsi} 
as a toy model which neatly interpolates between the standard Schwarzschild BH and a traversable WH in the  Morris-Thorne sense through the stage of a black bounce.\footnote{According to \cite{Simpson:2018tsi}, a black bounce is a minimum of the	spherical radius achieved in a nonstatic region of space-time beyond a BH horizon, in other words, it is a radial bounce in the Kantowski-Sachs homogeneous anisotropic metric in a BH interior. In \cite{Simpson:2018tsi}, an example of such behavior was presented as a toy model. Other examples of black bounces are	known in exact solutions of GR in the presence of phantom scalar fields \cite{Bronnikov:2005gm,Bronnikov:2006fu,Bolokhov:2012kn}, where such BHs were named ``black universes'' because black bounces were followed there by cosmological expansion and isotropization.} Its extension to axial symmetry was found in \cite{Mazza:2021rgq}. In the slow rotation approximation to linear order in the rotation parameter $\widetilde{\alpha}$, the SV metric is described in terms of a quasiglobal coordinate by the line element
\beq
ds^2=-\left( 1-\frac{2\,m}{\sqrt{r^2+a^2}}\right) dt^2+\frac{dr^2}{\left( 1-\frac{2\,m}{\sqrt{r^2+a^2}}\right)}-\frac{ 4m \widetilde{\alp}}{\sqrt{r^2+a^2}} \sin^2\theta dtd\phi+\left(r^2 +a^2\right) d\Omega^2\,.
\label{SV_met}
\eeq
The location of the throat is identified with the minimum of the circumferential radius $K(r)$, where one has $K'(r_0)=0 \Rightarrow r_0=0$. In the limit $a \rightarrow 0$, a slowly rotating Schwarzschild BH is recovered. When $a \neq 0$, we have different branches of the metric depending on the value of the dimensionless parameter $p\equiv a/m$. At $p=2$, there is a WH/BH threshold with the WH branch corresponding to $p >2$, while at 
$p \leqslant 2$ we have regular BH metrics. The radial profile of the metric function 
$f(r)\equiv -g_{tt}(r)$ is depicted in Fig.~\eqref{fig:SV_branches} for the three branches.
% -------------------------------------------------
% -------------------------------------------------

Under the radial coordinate redefinition $r \rightarrow \tilde{r} : \tilde{r}^2 = r^2+a^2$, the line element~\eqref{SV_met} can be recast to the MT frame where the metric has the following forms:
\beq
ds^2=-\left( 1-\frac{2\,m}{\tilde{r}}\right) dt^2+\left( 1-\frac{2\,m}{\tilde{r}}\right)^{-1} \left( 1-\frac{a^2}{\tilde{r}^2}\right)^{-1} d\tilde{r}^2-\frac{ 4m \widetilde{\alp}}{\tilde{r}} \sin^2\theta dtd\phi+\tilde{r}^2 d\Omega^2\,.
\label{SV_met_MT}
\eeq
The location of the WH throat $r_0$ in the MT frame is determined by the shape function satisfying $b(\tilde{r}_0)=\tilde{r}_0$. This condition is equivalent to identifying the roots of the equation
\beq
h(r) \equiv \left( 1-\frac{2\,m}{\tilde{r}}\right) \left( 1-\frac{a^2}{\tilde{r}^2}\right)=0\,.
\label{SV_MT_h(r)}
\eeq
Note that even though Eq.~\eqref{SV_MT_h(r)} has two roots $\tilde{r}=2m$ and $\tilde{r}=a$, the former is not a suitable choice for a WH throat because in this case $f(\tilde{r}_0)=0$, 
and an event horizon emerges. Thus the location of the throat has to be determined as 
$\tilde{r}_0=a$, and since $\tilde{r}$ is the curvature coordinate, the radius of the throat is identified with $\tilde{r}_0$. In this frame, the metric~\eqref{SV_met_MT} is parametrized 
exactly in terms of Eqs.~\eqref{RZ_DCC}--\eqref{WH_param_eq_B}, and the EPs have the following form:
\ba
\eps&=&\frac{2m}{a}-1\,,\quad a_0=0 \,,\quad f_0=1-\frac{2m}{a}\,,\quad a_i =0\quad \forall i \geqslant 1\,,\\
b_0&=&\frac{2m}{a}-1 \,,\quad h_0=0\,,\quad b_1=-\frac{2m}{a}\,,\quad b_i =0\quad \forall i \geqslant 2\,.
\ea

\vspace{0.5cm}
%=======================================================
\section{Parametrization of Wormholes in non-Morris-Thorne frames \label{Sec:V}}
%=======================================================

Thus far we have considered WH metrics that can be recast in the Morris-Thorne frame where the radial coordinate is the curvature coordinate and the throat is located at some positive value of the curvature coordinate $r_0 >0$. However, in some cases, such as in construction of numerical solutions, it might be convenient to obtain solutions in more general frames where the radial coordinate is not the curvature coordinate. In such cases the circumferential radius $K(r)$ is not identified with $r$ and consequently the location of the WH throat $r_0$ and its radius $K_0 \equiv K(r_0)$ are no longer identified. Furthermore, in the context of such solutions, the throat can be 
located at zero or even negative values of the radial coordinate. When $r_0=0$, the RZ proposal for the compact coordinate~\eqref{RZ_DCC} is no longer a suitable choice for a DCC upon which 
the parametrization can be constructed since it reduces to a constant $x=1$. To this end, the DCC part of the parametrization has to be appropriately modified in order to accommodate such cases. In the next subsection, we return to the SV metric as given in its original coordinate system~\eqref{SV_met}, prior to its transformation to the Morris-Thorne frame, in order to demonstrate how such a modification of the DCC might be realized.

% ---------------------------------------------------------------------
\subsection{Simpson-Visser wormhole revisited \label{Section:SV_revisited}}
% ---------------------------------------------------------------------
Let us begin by pointing out that since the SV metric~\eqref{SV_met} is symmetric relative to the throat, once we have the parametrization in $r \in [0,\infty)$, we at the same time have the one that is valid
for $r \in (-\infty,\infty)$. In the case of a reflection asymmetric metric, if one is interested in the region $r \leqslant r_0=0$, one can always perform two independent parametrizations for each side of the throat in a way that is analogous to the one we outline here for the $r \geqslant 0$ region.

Any dimensionless compact coordinate, which is to replace the RZ DCC given by \eqref{RZ_DCC}, must be constructed in such a way that it vanishes at the location of the throat $r_0=0$ and approaches unity asymptotically at spatial infinity. This is necessary for the DCC to map the whole region $r \in [r_0,\infty)$ to the compact range $x \in[0,1]$. To cover the parametrization of metrics in situations like the one described above, let us introduce the following class of optimized dimensionless compact coordinates $x(r)$, to be used instead of~\eqref{RZ_DCC} in the parametrization functions~\eqref{WH_param_fun_A},~\eqref{WH_param_fun_B} and~\eqref{WH_param_fun_C} when the throat is located at $r_0=0$:

\beq
x(r) \equiv 1-\left( \frac{R_0^{x_0}}{R_0^{x_0}+r^{x_0}} \right)^{\frac{x_{\infty}}{x_0}}\,,
\label{x_WH_R0}
\eeq
where $R_0$ is an arbitrary ``length-scale'' parameter which is introduced to make $x(r)$ dimensionless, and the parameters $x_{0}$ and $x_{\infty}$ in Eq.~\eqref{x_WH_R0} are defined via the asymptotics of the metric function we wish to parametrize (symbolically denoted here by $g(r)$)
\beq
x_{0} \equiv \lim_{r \rightarrow 0} \frac{g'(r)}{g(r)-g_0} r\,,
\label{x0_def}
\eeq
and
\beq
x_{\infty} \equiv \lim_{r \rightarrow \infty} \frac{g'(r)}{g_{\infty}-g(r)}r\,.
\label{x_inf_def}
\eeq

Since there are various ways to define the arbitrary parameter $R_0$, we will now discuss 
this feature of the optimized DCC~\eqref{x_WH_R0}. First of all, notice that in parametrization 
of BH metrics, the corresponding length scale $r_0$ that appears in the RZ DCC~\eqref{RZ_DCC}, must necessarily be identified with the size of the outer event horizon. This is required by the consistency of the parametrization equation~\eqref{RZ_param_eq_A} at the horizon, where $f(r_0)=0$, and so $x(r_0)=0$ must also be true. In the parametrization of WH metrics written 
in the MT frame, such as the cases we have considered in the previous section, 
it is the consistency of the parametrization equation~\eqref{WH_param_eq_B} that now demands that the parameter $r_0$ of the RZ DCC~\eqref{RZ_DCC} should be identified with the throat 
radius since in the MT frame one always has $h(r_0)=0$. Consequently, there is no freedom
(or ambiguity) in the definition of the length scale used in the construction of the RZ DCC.

On the other hand, the length-scale parameter $R_0$ in the optimized DCC of Eq.~\eqref{x_WH_R0} is not subject to any kind of ``consistency'' constraint in terms of the parametrization equations. Different choices for $R_0$ effectively correspond to different parametrizations, and this in turn  means that the possible choice of $R_0$ will affect the accuracy of parametrization for a given metric. To this end, when working with the optimized DCC, one should identify, on a case by case basis, the most convenient way of defining $R_0$ in terms of the parameters of the metric at hand in order to obtain the most accurate parametrization. In any case, as we will demonstrate, some of the most intuitive ways of choosing the length scale $R_0$ in a WH spacetime provide parametrizations 
that are very accurate even at the first order.

Returning to the SV metric, when written in the coordinate system of Eq.~\eqref{SV_met}, the optimized DCC~\eqref{x_WH_R0} that is suitable for the  parametrization of $f(r)$ is obtained 
via Eqs.~\eqref{x0_def} and~\eqref{x_inf_def} with $x_0=2$ and $x_\infty = 1$, respectively, 
and so it reads
\beq
x(r)=1-\sqrt{\frac{R_0^2}{R_0^2+r^2}}\,.     \label{SV_ODCC}
\eeq
By comparing both sides of the expansions of the parametrization equation~\eqref{WH_param_eq_A} at $x=0$ and at $x=1$ we determine the EP values. 
The first few are
\ba
\eps&=&\frac{2m}{R_0}-1 \,,\quad a_0=0  \,,\quad f_0= 1-\frac{2m}{a} \,,\quad a_1 = 2 \eps+3 f_0 -1+\frac{2 m R_0^2}{a^3}\,,\\
a_2 &=&\frac{3}{2 a_1}\left( 4 \eps +5 f_0 -3 a_1 -1 +\frac{2m R_0^4}{a^5}\right)\,.
\ea
We now turn to the definition of the length parameter $R_0$ for the SV metric. Perhaps the most natural and intuitive way to define a length scale associated with a WH geometry is given by identifying $R_0$ with the radius of the WH throat, 
\beq
R_0\equiv K(r_0)=a\,.\label{SV_ODCC_R0_1}
\eeq
For the SV metric, this choice provides an exact parametrization in all orders in the expansion. This, however, cannot be expected to hold true for any arbitrary metric function and optimized DCC, and one should, in principle, test different choices for the length scale parameter $R_0$ constructed by the parameters of the solution at hand in order to obtain the optimal accuracy. In what follows we propose two more ways to construct length-scale parameters $R_0$ from the metric in a systematic way.

Since we are interested in the parametrization of asymptotically flat geometries, the metric functions asymptote to some finite value (not necessarily the same) at both infinities. Furthermore, they must be finite everywhere, and consequently (if we exclude the trivial case where the metric function is constant) we have the emergence of at least one inflection point. Then one may identify $R_0$ with the value of the circumferential radius $K(r)$ at the location of the inflection point $r_{\rm inf}$ of the metric function. For the SV WH this condition yields
\beq
f''(r_{\rm inf})=0 \rightarrow R_0 \equiv K(r_{\rm inf}) =
a \sqrt{\frac{3}{2}}\,.
\label{SV_ODCC_R0_ip}
\eeq
Another definition of the length scale $R_0$ in terms of the free parameters of the metric in a systematic way can be via the value of the second derivative of the metric function 
evaluated at the throat. In our example we find
\beq
R_0 \equiv \frac{1}{ \sqrt{f''(r_0)}}=\sqrt{\frac{a^3}{2m}}\,,  \label{SV_ODCC_R0_d2th}
\eeq
while in cases where $f''(r_0)=0$, the above definition could be replaced by $R_0=f'(r_0)^{-1}$. In any case, the aforementioned definitions for $R_0$ do not, by any means, exhaust  all possible
ways to define $R_0$ in terms of the free parameters of a metric, and when employing the approach of the optimized DCC, one must inevitably try different values for $R_0$.

In Tables~\eqref{table:SV_R0_ip} and~\eqref{table:SV_R0_d2th} with $R_0$ defined as in  Eq.~\eqref{SV_ODCC_R0_ip} and in Eq.~\eqref{SV_ODCC_R0_d2th}, respectively, we present the accuracy of the lowest orders of the approximation of the metric function 
$f(r)$ for various values of the parameter $p \equiv a/m$.

\begin{table}[ht!]
\center{
\caption{The percentage of maximum absolute relative error for various approximation 
orders and values of the parameter $p \equiv a/m$. Here, the length scale parameter 
$R_0 = a \sqrt{3/2}$ has been identified with the value of the circumferential radius at location of the inflection point of the metric function $f(r)$ as in Eq.~\eqref{SV_ODCC_R0_ip}.}
\begin{tabular}{|c|c|c|c|c|c|c|}
 \hline
 order
  &  $p=2.01$ & $p=2.1$ & $p=2.4$ & $p=2.5$ & $p=2.7$ & $p=2.99$ \\
 \hline\hline
  1 & 1.10913 & 0.97295 & 0.70486 & 0.64755 & 0.55808 & 0.46607\\
 \hline
  2 &  0.04582 & 0.04170 & 0.03230 & 0.03008 & 0.02647 & 0.02257 \\
 \hline
  3 &  0.00974 & 0.00899 & 0.00719 & 0.00674 & 0.00600 & 0.00518  \\
 \hline
   4 & 0.00482 & 0.00450 & 0.00370 & 0.00349 & 0.00314 & 0.00275 \\
 \hline
\end{tabular}
\label{table:SV_R0_ip}}
\end{table}

\begin{table}[ht!]
\center{
\caption{The percentage of maximum absolute relative error for various approximation orders 
and values of the parameter $p \equiv a/m$. Here, the length scale parameter 
$R_0=a^{3/2}/\sqrt{2m}$ has been identified in terms of the value of the second-order 
derivative of the metric function $f(r)$ at the location of the throat, as in Eq.~\eqref{SV_ODCC_R0_d2th}.}
\begin{tabular}{|c|c|c|c|c|c|c|}
 \hline
 order
  &  $p=2.01$ & $p=2.1$ & $p=2.4$ & $p=2.5$ & $p=2.7$ & $p=2.99$ \\
 \hline\hline
  1 & 0.00010 & 0.00872 & 0.10724 & 0.15608 & 0.26903 & 0.45664  \\
 \hline
  2 &  0.00001 & 0.00107 & 0.00938 & 0.01229 & 0.01731 & 0.02233 \\
 \hline
  3 &  $\mathcal{O}(10^{-7})$ & 0.00004 & 0.00106 & 0.00162 & 0.00293 & 0.00508  \\
 \hline
   4 & $\mathcal{O}(10^{-11})$ & $\mathcal{O}(10^{-7})$ & 0.00007 & 0.00017 & 0.00062 & 0.00261
   \\
 \hline
\end{tabular}
\label{table:SV_R0_d2th}}
\end{table}

For both choices of $R_0$ we see quick convergence and excellent accuracy of the 
approximation already at the first order, even as the WH/BH threshold $(p=2)$ is approached. Furthermore, a qualitative difference between the two parametrizations is also evident. 
When $R_0$ is defined as in Eq.~\eqref{SV_ODCC_R0_ip}, we see in Table~\eqref{table:SV_R0_ip} that the parametrization becomes less accurate as the 
WH/BH threshold is approached, which is not so for the parametrization with $R_0$ 
defined as in Eq.~\eqref{SV_ODCC_R0_d2th} according to the entries of Table~\eqref{table:SV_R0_d2th}. This verifies that different choices of the length-scale 
parameter might provide an accuracy in some regions of the parameter space better than in
others. To this end, one is encouraged to try different definitions of $R_0$ in order to identify 
the one that yields the optimal accuracy in the parameter range one is interested in.

In terms of the quasiglobal coordinate of the frame~\eqref{SV_met}, the $g_{t\phi}(r,\theta)$ 
metric component of the slowly rotating SV metric does not explicitly correspond to the 
Lense-Thirring term as in its MT frame representation~\eqref{SV_met_MT}. Thus a 
parametrization by means of Eqs.~\eqref{WH_param_eq_C} and~\eqref{WH_param_fun_C} 
with the optimized DCC of Eq.~\eqref{SV_ODCC} may be performed. A few first EPs 
in this case turn out to be
\beq
c_0=\frac{4m \widetilde{\alp}}{R_0} \,,\quad g_0= -\frac{4m \widetilde{\alp}}{a}\,,\quad c_1 = c_0+2 g_0+\frac{4 m R_0^2 \widetilde{\alp}}{a^3} \,.
\eeq
If, according to Eq.~\eqref{SV_ODCC_R0_1}, we choose to identify the length-scale parameter
with the radius of the throat, i.e., $R_0=a$, it turns out that $c_1=0$, then the parametrization function~\eqref{WH_param_fun_C} reproduces the Lense-Thirring term in terms of the compact coordinate exactly.

\begin{figure*}[ht!]
%\centering
\includegraphics[width=0.5\linewidth]{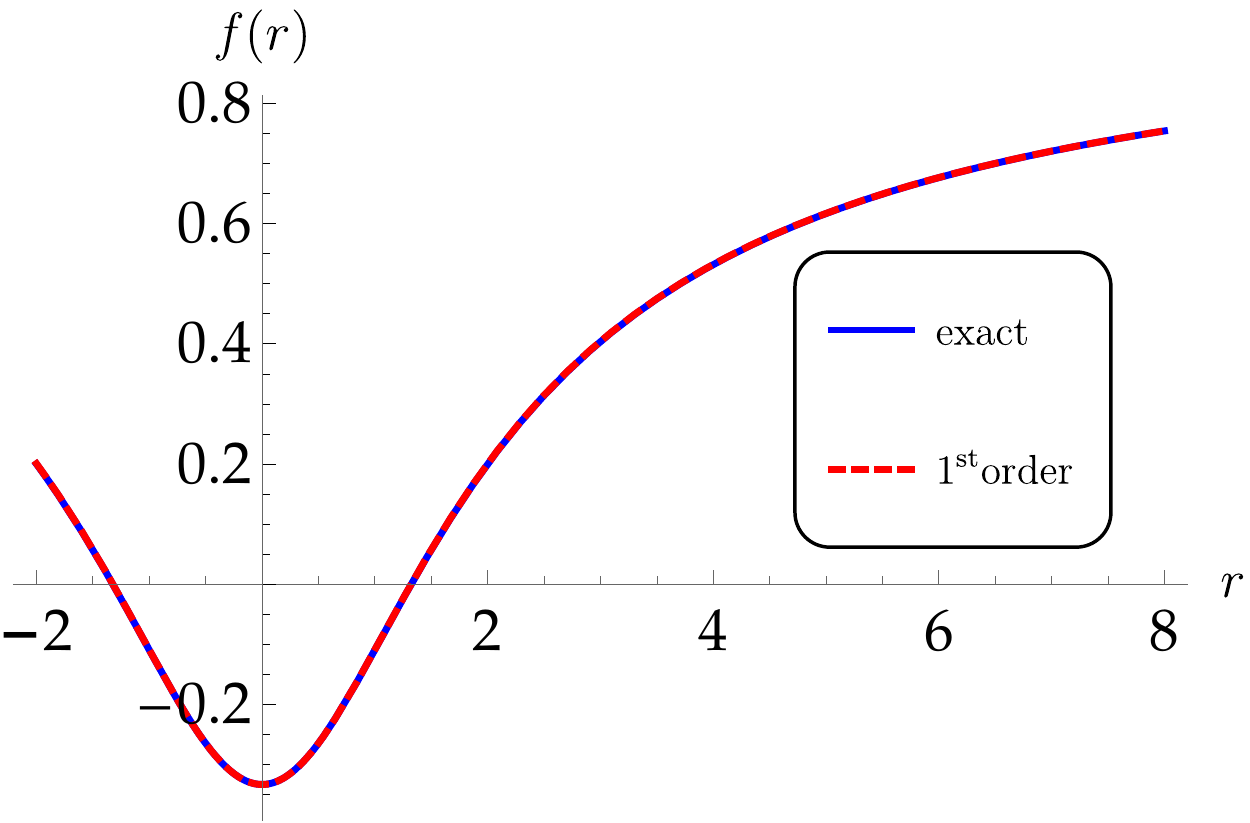}
\includegraphics[width=0.5\linewidth]{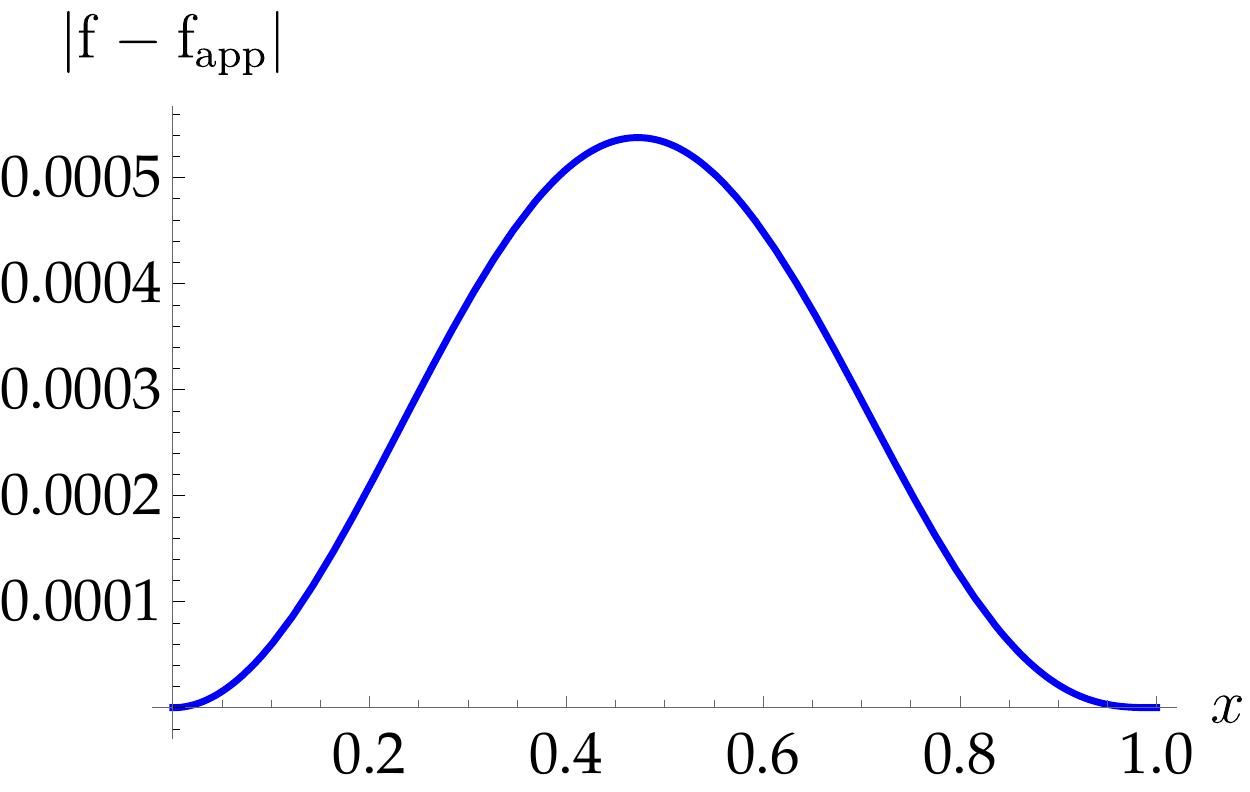}
\caption{The regular black hole branch, for $p \equiv a/m =1.5$. 
Left panel: the exact metric function  $f(r) \equiv -g_{tt}(r)$~\eqref{SV_met} (blue curve) and 
its first-order approximation $f_{\rm app}(r)$ (red dashed curve). Right panel: their absolute difference in terms of the optimized dimensionless compact coordinate $x \in [0,1]$~\eqref{SV_ODCC} for $R_0$ as given in Eq.~\eqref{SV_ODCC_R0_d2th}.
}
\label{fig:SV_Regular_BH}
\end{figure*}

Finally, we mention that our parametrizations (in both frames ~\eqref{SV_met},\eqref{SV_met_MT} and for both choices of $R_0$) can also be very successfully applied for the regular BH branch of the metric. Using the parametrization approach based on the optimized DCC \eqref{SV_ODCC} on the left panel of Fig.~\eqref{fig:SV_Regular_BH}, we plot the exact metric function $f$ in the regular BH branch ($p=1.5$) and its first-order approximation $f_{\rm app}$ in terms of the quasiglobal coordinate. On the right panel of the same figure, we plot the absolute difference between $f$ and $f_{\rm app}$ in terms of the optimized DCC.

% ----------------------------------------------------------------------
\subsection{Wormhole branch of the anti-Fisher solution}
% ----------------------------------------------------------------------

\noindent 
General relativity with a massless minimally coupled scalar field allows for a solution containing a naked singularity, known as the Fisher solution 
\cite{Fisher:1948yn}. When the kinetic term has the opposite sign, a solution emerges
which has a WH branch and was called by one of us the ``anti-Fisher solution''
by analogy with the ``anti-de Sitter solution'' \cite{Bronnikov:1973fh}. The metric 
in the WH branch of the anti-Fisher solution can be written as  
\cite{Bronnikov:1973fh,Kashargin:2007mm}
\beq
ds^2=-e^{2u(r)}dt^2+e^{-2u(r)} \left[ dr^2+\left(r^2+a^2 \right)d\Omega^2 \right]\,,\qquad u(r) \equiv \frac{m}{a}\left(\arctan\frac{r}{a}-\frac{\pi}{2} \right)\,,
\label{anti-Fisher}
\eeq
where we assume that the arbitrary parameters satisfy $m \geqslant 0$ and $a >0$. In the case of the line element above, we are forced to perform a parametrization in the coordinate frame with 
$K(r) \neq r$ since we cannot analytically move to the MT frame \eqref{MT_ansatz_le}. This means that, strictly speaking,  a different radial coordinate is 
used here, and the parametrization in this case is different from the ones discussed in Sec.~\ref{Sec:IV} as was the case with the SV metric in the frame of the previous subsection. On the 
other hand, when comparing various parametrized WH geometries, the coordinate 
choice must be unique.  Therefore, this example provides further verification that our proposed parametrization
can also provide an effective description of WH geometries in other coordinate systems beyond the ones employing the curvature coordinate.

Notice that in the coordinate system~\eqref{anti-Fisher} we have $f(r)=h(r)=e^{2u(r)}$, and so $r$ is a 
quasiglobal but not curvature coordinate according to the classification of Sec.~\ref{Sec:II}. 
Minimization of $K^2(r)=e^{-2u(r)}\left(r^2+a^2 \right)$ reveals that the throat is located 
at $r_0=m$ and has a radius
\beq
K_0 \equiv K(r_0) = e^{-u(m) }\sqrt{m^2+a^2}\,.
\eeq
If one is not interested in the limit $m \rightarrow 0$ (which will be discussed separately), we can be constrained by the case $r_0 =m \neq 0$. In the latter case we are going to employ the RZ DCC~\eqref{RZ_DCC} for our parametrization, keeping always in mind that the radial coordinate is not the curvature coordinate and consequently the parametrization here is conceptually distinct from the cases considered in the Morris-Thorne frame in Sec.~\ref{Sec:IV}. The parametrization of $f(r)$ is then obtained in terms of Eqs.~\eqref{WH_param_fun_A} and~\eqref{WH_param_eq_A}. The expansions of the parametrization equation~\eqref{WH_param_eq_A} at the boundaries of the parametrization range $r \in [r_0,\infty) \rightarrow x \in [0,1]$ determine the EPs, 
the first few of which have the following expressions:
\ba
\eps&=&1\,,\quad a_0=2 \,,\quad f_0=e^{2 u(m)}\,,\quad a_1 =f_0 \left( 3+\frac{2 m^2}{a^2+m^2} \right)-1\,, \\
a_2 &=& \frac{6f_0-4 a_1-2}{a_1}\,,\quad a_3 = -\frac{f_0 \left[ \frac{2 m^4}{3 \left( a^2+m^2 \right)^2} -10 \right]+ a_1 \left[10+a_2 \left(5+a_2 \right)\right]+4}{ a_1 a_2} \,.
\label{anti-Fisher_EPs}
\ea
Upon substituting the EPs above into the expression of the parametrization function~\eqref{WH_param_fun_A}, we can compute the absolute relative error (ARE) and MARE for the approximation of the anti-Fisher solution at various orders in the continued-fraction expansion.

In order to test the convergence of the approximation, in Table~\eqref{table:anti-Fisher} we give  the percentage of MARE for the first four orders in the approximation and for various values of the dimensionless parameter $p \equiv a/m$. It is evident that for all values of $p$ the MARE is decreased when the order of approximation is growing.

\begin{table}[H]
\center{
\caption{The percentage of maximum absolute relative error between the exact metric function $f(r)$ as given in Eq.~\eqref{anti-Fisher} and its lowest-order approximations for various values of the dimensionless parameter $p \equiv a/m$.}
\begin{tabular}{|c|c|c|c|c|c|c|c|}
 \hline
 order
  &  $p=0.01$  & $p=0.2$ & $p=0.3$ & $p=0.4$ & $p=0.5$ & $p=0.6$  \\
 \hline\hline
  1 & 1.45307 & 1.32808 & 1.17244 & 0.95673 & 0.68424 & 0.36334 \\
 \hline
  2 & 0.02816 & 0.03420 & 0.04403 & 0.06250 & 0.09516 & 0.14985 \\
 \hline
  3 & 0.00308 & 0.00392 & 0.00575 & 0.01068 & 0.02378 & 0.05888  \\
 \hline
   4 & 0.00013 & 0.00022 & 0.00040 & 0.00075 & 0.00134 & 0.00225  \\
 \hline
\end{tabular}
\label{table:anti-Fisher}
}
\end{table}

\noindent At this point we make a remark about the limit $m \rightarrow 0$ of our parametrization for which the anti-Fisher WH~\eqref{anti-Fisher} reduces to the Ellis-Bronnikov WH \cite{Ellis:1973yv,Bronnikov:1973fh}
\beq
ds^2=-dt^2+dr^2+\left(r^2+a^2 \right) d\Omega^2\,.
\eeq
At first glance, someone might consider the use of the RZ DCC as a naive choice for the compact coordinate given the fact that the length scale parameter $r_0=m$ vanishes in that limit and consequently we are seemingly dealing with a situation that has been remedied with the optimization of the DCC in the previous section. However, the two cases are radically different since in the case of the SV metric one has $r_0=0$ fixed throughout the parametric space while here, this value is only obtained in a certain limit. When $r_0=m \rightarrow0$ the RZ DCC, as we have already mentioned, approaches asymptotically a constant $x=1$ and this causes the parametrization equation~\eqref{WH_param_eq_A} to reduce to its asymptotic value at infinity $A(1)=1$. At the same time, all the EPs~\eqref{anti-Fisher_EPs} (including the higher-order ones) in our parametrization of the anti-Fisher metric are finite constants for $m=0$ and $a>0$ and consequently the parametrization reduces smoothly to the Ellis-Bronnikov limit where $\lim_{m \rightarrow 0}f(r) \rightarrow 1$.

Next, we are going to parametrize the metric of a slowly rotating anti-Fisher WH. In \cite{Kashargin:2007mm}, the first-order corrected metric with respect to the rotation 
parameter $\widetilde{\alp}$ was found, and it is given by
\beq
ds^2_{\widetilde{\alp}}=ds^2-2\widetilde{\alp} \omega^{(1)} e^{-2 u(r)} \left(r^2+a^2 \right) \sin^2 \theta dt d\phi \,,\qquad u(r) \equiv \frac{m}{a}\left(\arctan\frac{r}{a}-\frac{\pi}{2} \right)\,,
\eeq
where $ds^2$ is the line element of the static anti-Fisher WH as given in Eq.~\eqref{anti-Fisher}. Here the angular velocity metric function is 
\beq
\omega(r) \equiv \widetilde{\alp} \omega^{(1)}=\frac{\widetilde{\alp} \left(1-e^{4 u(r)}\right)\left[ 1+\frac{4 m \left( r+2m \right)}{r^2+a^2}\right] }{a \left[ 1- e^{-2 \pi m/a} \left(1+8m^2 /a^2 \right) \right]}\,.
\eeq
The parametrization for the metric function
\beq
g(r)\equiv-2\widetilde{\alp} \omega^{(1)} e^{-2 u(r)} \left(r^2+a^2 \right)\,,
\label{SR_anti-Fisher_g(r)}
\eeq
is obtained in terms of  Eqs.~\eqref{WH_param_fun_C} and~\eqref{WH_param_eq_C}. The asymptotic parameter $c_0$ and the first few  near-field parameters are given by 
\ba
c_0&=& \frac{16 a  e^{\frac{2 \pi  m}{a}} \left(a^2+4 m^2\right) \widetilde{\alp}}{3 \left[a^2 \left(e^{\frac{2 \pi  m}{a}}-1\right)-8 m^2\right]}\,,g_0=\frac{3\, c_0\, e^{2 u(m)} \left[ a^2+13 m^2-e^{-4 u(m)} \left( a^2 +m^2 \right)\right]}{8 \left( a^2+4m^2 \right)}\,,
\\
c_1&=&c_0 \left[1+\, \frac{2\,g_0}{c_0}+\frac{3 m^2 e^{2 u(m)}}{a^2+m^2} \right]\,,\quad c_2=\frac{2\,c_0}{c_1}+\frac{g_0}{c_1} \left(3-\frac{m^2}{a^2+m^2} \right)-3\,.
\ea
Notice that, as should be the case, the value of $c_0$ above is exactly equal to the expression 
$4J/r_0=4J/m$, where $J$ is given by Eq.~(26) of \cite{Kashargin:2007mm}.

Next, in order to test the convergence of the parametrization, in Table~\eqref{table:anti-Fisher_SR} we give the MARE for various values of the parameter $p\equiv a/m$ at the first four orders in the approximation of~\eqref{SR_anti-Fisher_g(r)}. Indeed, we find that the error decreases as more higher-order terms are taken into account, and this verifies the convergence of the approximation.

\begin{table}[H]
\center{
\caption{The percentage of maximum absolute relative error between the exact metric function $g(r)/\widetilde{\alp}$ as given in Eq.~\eqref{SR_anti-Fisher_g(r)} and its lowest-order approximations for various values of the dimensionless parameter $p \equiv a/m$.}
\begin{tabular}{|c|c|c|c|c|c|c|c|}
 \hline
 order
  &  $p=0.01$  & $p=0.2$ & $p=0.3$ & $p=0.4$ & $p=0.5$ & $p=0.6$  \\
 \hline\hline
  1 & 2.98055 &	3.08447 & 3.20575 &	3.36015 &	3.53514 & 3.71807 \\
 \hline
  2 & 0.30280 & 0.12709 & 0.08816 & 0.34524 & 0.63398 & 0.92509 \\
 \hline
  3 & 0.26435 &	0.15030 & 0.09399 & 0.08777 & 0.13282 & 0.21895  \\
 \hline
   4 & 0.04350 & 0.04729 & 0.08854 & 0.00375 & 0.00079 & 0.01529  \\
 \hline
\end{tabular}
\label{table:anti-Fisher_SR}
}
\end{table}

\vspace{0.5cm}
%=======================================================
\section{Gauge-invariant tests of the parametrization accuracy \label{Sec:VΙ}}
%=======================================================

In this section, we turn to computing the radii of shadows and quasi-normal modes 
as gauge-invariant tests of the parametrization accuracy.

%==========================
\subsection{Wormhole shadows}
 
\noindent Here, we give a brief overview of the formalism suggested in \cite{Synge:1966okc} 
and generalized in \cite{Perlick:2015vta} for computation of the shadow radius of an arbitrary 
static, spherically symmetric compact object, including WH solutions that are of interest to us 
in this work. Consider the general ansatz for a static and spherically symmetric geometry,
\beq
ds^2=-f(r)dt^2+\frac{dr^2}{h(r)}+K^2(r)d\Omega^2\,.
\label{le_general_shadows}
\eeq
As is pointed out it in \cite{Perlick:2015vta}, it is possible to derive an expression for the radius 
of the shadow of an arbitrary compact object in terms of the general metric functions that 
appear in Eq.~\eqref{le_general_shadows}, and consequently it may be applied to both 
BH and WH metrics. The shadow of a compact object is closely related to the location of 
its photon sphere which we shall denote here by $r_{\rm ph}$. It is convenient to introduce the function
\beq
w^2(r) \equiv \frac{K^2(r)}{f(r)}\,,
\eeq
in terms of which $r_{\rm ph}$ is determined as the solution to the equation
\beq
\frac{d w^2(r_{\rm ph})}{dr}=0\,.     \label{uph_condition}
\eeq
Next, the angular radius of the shadow, as seen by a distant static observer located at 
$r_{\rm O}$, is obtained via
\beq
\sin^2{a_{\rm sh}} = \frac{w^2(r_{\rm ph})}{w^2(r_{\rm O})}\,.
\eeq
Finally, under the assumption that the observer is located sufficiently far from the compact 
object, i.e., $r_{ \rm O} \gg r_0$, where $r_0$ is a characteristic length scale that can be 
identified with the radius of the WH throat, the following conditions are satisfied:
\beq
f(r_{\rm O}) \simeq 1\,,\quad K^2(r_{\rm O}) \simeq r_O^2\,,
\eeq
and one finds that the radius of the shadow is given by
\beq
R_{\rm sh} \simeq r_{\rm O} \sin{a_{\rm sh}} 
\simeq w(r_{\rm ph})= \frac{K(r_{\rm ph})}{\sqrt{f(r_{\rm ph})}}\,.         \label{Rsh}
\eeq

To test the accuracy of the metric approximations of the previous sections, in each case  we are now going to compute the value of the shadow radius~\eqref{Rsh} using the  exact expressions for the metric functions and compare its value to the one obtained  when employing the corresponding approximations of the metric at various orders of  the expansion. As is evident from Eq.~\eqref{Rsh}, the shadow radius is completely  specified in terms of the metric function $f(r)$ and the circumferential radius $K(r)$  and is thus independent of the form of $h(r)$. Thus, in the next section we will use the  anti-Fisher~\eqref{anti-Fisher} and Simpson-Visser~\eqref{SV_met} metrics as examples for which $f(r)$ is not exactly parametrizable.

% -------------------------------------------------------------
\subsection{Shadows of the anti-Fisher wormhole}
% -------------------------------------------------------------

\noindent
The anti-Fisher solution~\eqref{anti-Fisher} is written in terms of the quasi-global coordinate and the two relevant metric functions have the following form:
\beq
f(r)=e^{2u(r)}\,,\quad K^2(r)=e^{-2u(r)} \left(r^2+a^2 \right)\,,\quad u(r) \equiv \frac{m}{a}\left(\arctan\frac{r}{a}-\frac{\pi}{2} \right)\,.
\eeq
Thus we have
\beq
w^2(r)=e^{-4 u(r)} \left(r^2+a^2\right)\,,
\eeq
and the condition $d (w^2(r)/dr=0)$ yields the photon sphere radius $r_{\rm ph}=2m$. Then via Eq.~\eqref{Rsh} we obtain the exact expression for the shadow radius
\beq
R_{\rm sh}= e^{-2\, u(r_{\rm ph})} \sqrt{4 m^2 + a^2 }\,.
\label{Rsh_anti-F}
\eeq
In terms of the dimensionless parameter $p \equiv a/m$, we have compared the value obtained by Eq.~\eqref{Rsh_anti-F} with the one obtained via the approximations of the metric~\eqref{anti-Fisher} at various approximation orders, and the results are given in Table~\eqref{table:anti-Fisher_Rsh}. We point out that the value of $r_{\rm ph}$, which we used to obtain the approximate values for the shadow radius, is obtained for each order in the approximation via 
extremization of the function $w^2(r)$.

\begin{table}[H]
\center{
\caption{The percentage of absolute relative error between the exact value of the shadow radius~\eqref{Rsh_anti-F} and its value as obtained via Eq.~\eqref{Rsh} for various approximation orders of the metric function $f(r)$. The dimensionless parameter $p \equiv a/m$. }
\begin{tabular}{|c|c|c|c|c|c|c|c|}
 \hline
 order
  &  $p=0.01$  & $p=0.2$ & $p=0.3$ & $p=0.4$ & $p=0.5$ & $p=0.6$  \\
 \hline\hline
  1 & 1.04775 & 0.95530 & 0.83863 & 0.67406 & 0.46096 & 0.19918 \\
 \hline
  2 & 0.02753 & 0.03339 & 0.04295 & 0.06093 & 0.09281 & 0.14633 \\
 \hline
  3 & 0.00304 & 0.00387 & 0.00570 & 0.01061 & 0.02367 & 0.05883  \\
 \hline
   4 & 0.00011 & 0.00019 & 0.00034 & 0.00063 & 0.00112 & 0.00186 \\
 \hline
\end{tabular}
\label{table:anti-Fisher_Rsh}
}
\end{table}

% ------------------------------------------------------------------------
\subsection{Shadows of the Simpson-Visser wormhole}
% ------------------------------------------------------------------------

\noindent
The two metric functions which are relevant to the computation of shadows from the line element of Eq.~\eqref{SV_met} are the following:
\beq
f(r)=\left( 1-\frac{2\,m}{\sqrt{r^2+a^2}}\right)\,,\quad K^2(r)=\left(r^2 +a^2\right)\,.
\eeq
Thus we have
\beq
w^2(r)=\frac{\left(r^2+a^2 \right)}{\left( 1-\frac{2\,m}{\sqrt{r^2+a^2}}\right)}\,,
\eeq
and the photon sphere radius corresponds to $r_{\rm ph}=\sqrt{(3m)^2-a^2}$. This means that the SV WH branch of the metric has a photon sphere when $ 2 m  \leqslant a \leqslant 3 m $ with the WH/BH threshold corresponding to $a = 2m$. Then via Eq.~\eqref{Rsh} we obtain the exact expression for the shadow radius
\beq
R_{\rm sh}=3 \sqrt{3}\, m\,,
\label{Rsh_SV}
\eeq
where we can see that it does not depend on the parameter $a$ and it is  identified with the shadow radius of the Schwarzschild BH. For a detailed discussion on the degeneracy of shadows in the SV spacetime see \cite{Junior:2021atr}. For various values of the dimensionless parameter $p \equiv a/m $, we have compared the value obtained by Eq.~\eqref{Rsh_SV} with the one obtained via the approximations of the metric at various orders in the approximation. The obtained percentage of the absolute relative error are given in Table~\eqref{SV_ODCC_R0_ip_Rsh} for $R_0$ as defined in  Eq.~\eqref{SV_ODCC_R0_ip} and in Table~\eqref{SV_ODCC_R0_d2th_Rsh} for $R_0$ as defined in Eq.~\eqref{SV_ODCC_R0_d2th}.

\begin{table}[H]
\center{
\caption{The percentage of absolute relative error between the analytic value of the shadow radius~\eqref{Rsh_SV} of the Simpson-Visser wormhole and its value as obtained via Eq.~\eqref{Rsh} for various approximation orders of the metric. The WH/BH threshold corresponds to $p=2$, while for $p>3$ the WH has no photon sphere. Here, we have used 
$R_0 = a\sqrt{3/2}$ according to Eq.~\eqref{SV_ODCC_R0_ip}.}
\begin{tabular}{|c|c|c|c|c|c|c|}
 \hline
 order
  &  $p=2.01$ & $p=2.1$ & $p=2.4$ & $p=2.5$ & $p=2.7$ & $p=2.99$ \\
 \hline\hline
  1 & 0.54727 & 0.47968 & 0.25291 & 0.18463 & 0.07285 & 0.00009 \\
 \hline
  2 & 0.01973 & 0.01568 & 0.00544 & 0.00329 & 0.00076 & $ \mathcal{O}(10^{-8})$ \\
 \hline
  3 & 0.00278 & 0.00200 & 0.00046 & 0.00023 &  0.00003 & $\mathcal{O}(10^{-11})$  \\
 \hline
   4 & 0.00060 & 0.00039 & 0.00005 & 0.00002 & $\mathcal{O}(10^{-6})$ & $\mathcal{O}(10^{-13})$  \\
 \hline
\end{tabular}
\label{SV_ODCC_R0_ip_Rsh}}
\end{table}

\begin{table}[H]
\center{
\caption{The percentage of absolute relative error between the analytic value of the shadow radius~\eqref{Rsh_SV} of the Simpson-Vissser wormhole and its value as obtained via Eq.~\eqref{Rsh} for various approximation orders of the metric. The WH/BH threshold corresponds to $p=2$ while for $p>3$ the WH has no photon sphere.  Here, we have used  $R_0=a^{3/2}/\sqrt{2m}$ according to Eq.~\eqref{SV_ODCC_R0_d2th}.}
\begin{tabular}{|c|c|c|c|c|c|c|}
 \hline
 order
  &  $p=2.01$ & $p=2.1$ & $p=2.4$ & $p=2.5$ & $p=2.7$ & $p=2.99$ \\
 \hline\hline
  1 & 0.00005 & 0.00435 & 0.04207 & 0.04878 & 0.03771 & 0.00009 \\
 \hline
  2 &   $\mathcal{O}(10^{-5})$
  & 0.00047  & 0.00197  & 0.00166 & 0.00059 &  $\mathcal{O}(10^{-8})$ \\
 \hline
  3 &   $\mathcal{O}(10^{-8})$ & 0.00001 & 0.00008 & 0.00007 & 0.00002 &  $\mathcal{O}(10^{-11})$  \\
 \hline
   4 &  $\mathcal{O}(10^{-11})$ &  $\mathcal{O}(10^{-7})$ &  $\mathcal{O}(10^{-6})$ &  $\mathcal{O}(10^{-6})$ &  $\mathcal{O}(10^{-6})$ &  $\mathcal{O}(10^{-13})$  \\
 \hline
\end{tabular}
\label{SV_ODCC_R0_d2th_Rsh}}
\end{table}

\begin{figure*}[ht!]
%\centering
\includegraphics[width=0.51\linewidth]{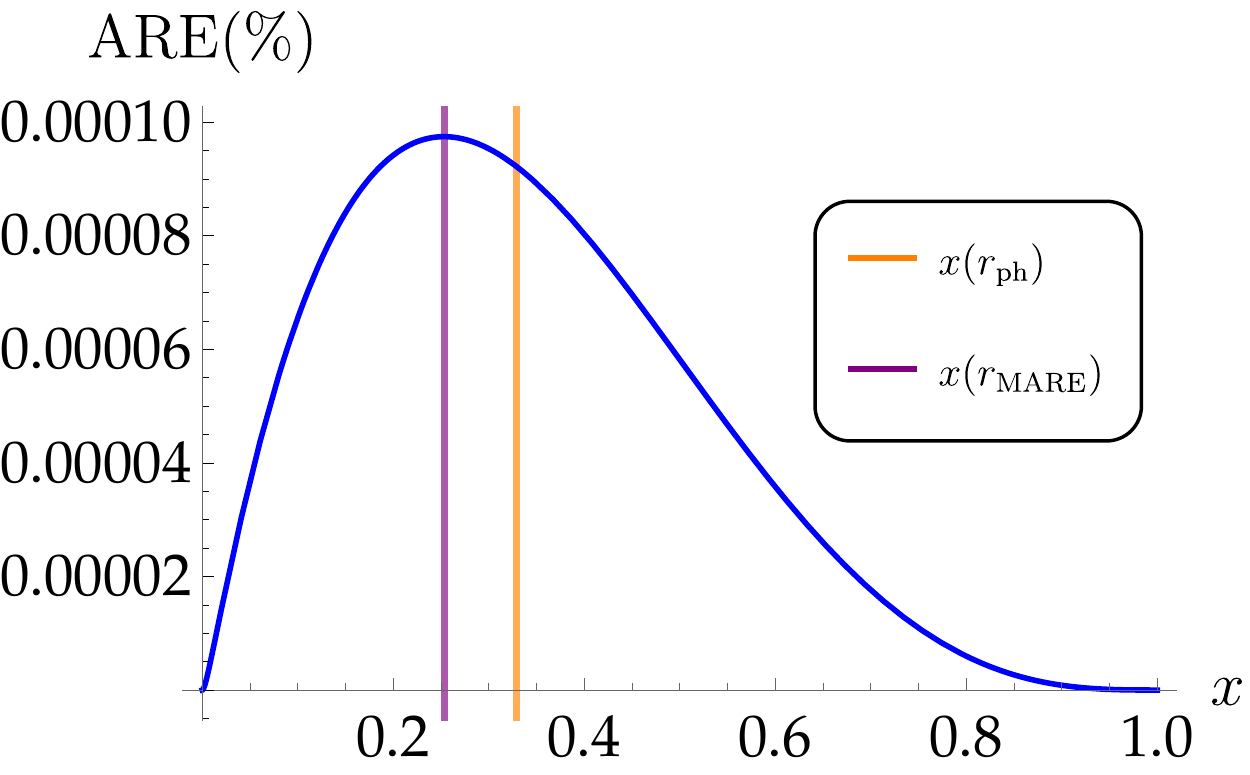}
\includegraphics[width=0.49\linewidth]{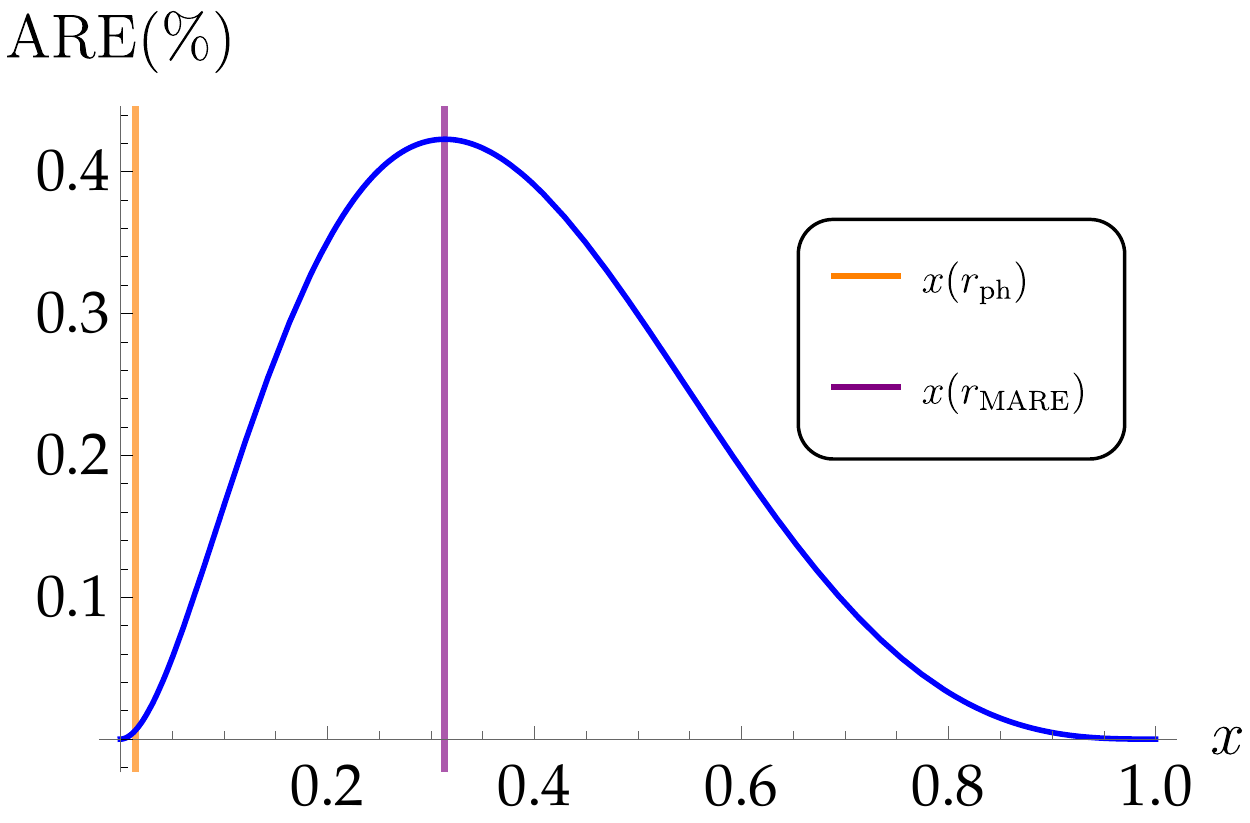}
\caption{The percentage of the absolute relative error of the first-order approximation of 
the metric in terms of the optimized compact coordinate $x \in [0,1]$~\eqref{SV_ODCC} 
with $R_0$ as given in Eq.~\eqref{SV_ODCC_R0_d2th}. Also depicted are the location 
of the MARE $r_{\rm MARE}$ (purple line) and of the photon sphere $r_{\rm ph}$ (orange line) 
as obtained with the first-order approximation of the metric. Left panel: near 
the wormhole/black hole threshold with $p \equiv a/m =2.01$. Right panel: near the limit 
where the wormhole still has a photon sphere, with $p=2.94$.}
\label{fig:SV_MARE_and_rph}
\end{figure*}

First, notice that in both Tables~\eqref{SV_ODCC_R0_ip_Rsh} and~\eqref{SV_ODCC_R0_d2th_Rsh}, the accuracy in the approximation of the shadow 
radius is in all cases better than the accuracy of the corresponding order of the approximation 
for the metric as given in Tables~\eqref{table:SV_R0_ip} and~\eqref{table:SV_R0_d2th}, respectively. This is a consequence of the fact that the location of the photon sphere radius 
$r_{\rm ph}$, which is important for the calculation of the shadow radius, is not identified with the location of the MARE of the metric $r_{\rm MARE}$, as can be seen in Fig.~\eqref{fig:SV_MARE_and_rph}. Close to the WH/BH threshold, i.e., in the limit 
$p \rightarrow 2$, the photon sphere lies in the region $r_{\rm ph} > r_{\rm MARE}$. Then, gradually, as the value of $p$ increases, it attains a critical value for which $r_{\rm ph}$ is 
identified with $r_{\rm MARE}$ and consequently, as $p$ further increases, we end up 
with $r_{\rm ph} < r_{\rm MARE}$.

Second, notice that according to the entries of Tables~\eqref{SV_ODCC_R0_ip_Rsh} and~\eqref{SV_ODCC_R0_d2th_Rsh}, close to the limit where the WH has a photon
sphere ($p=3$), the error in the shadow radius is highly suppressed in both cases. 
This is attributed to the fact that in that limit $r_{\rm ph}$ shifts towards the lower boundary
of the parametrization region ($x=0$) where the throat is located, see the right panel of Fig.~\eqref{fig:SV_MARE_and_rph}, and as we have already discussed, the error between the exact metric and its approximation at any order is exactly zero at both $x=0$ and $x=1$.

In conclusion, independently of the choice of the length scale parameter $R_0$ of the parametrization, for a given approximation order, the obtained values of the shadow radius 
error at some $p$ can be understood as an interplay between the following criteria:
The relative location of $r_{\rm ph}$ w.r.t. $r_{\rm MARE}$, the value of MARE at that $p$
and, finally, how far is  $r_{\rm ph}$ located from the throat.

% ---------------------------------------------
\subsection{Quasinormal modes}
% ---------------------------------------------

Here, as another example of observable quantities, we will calculate the fundamental
quasinormal modes of the electromagnetic field propagating in a wormhole background. 
Quasinormal modes are characteristic frequencies of a compact object (be it a black hole or a wormhole) which are independent of the initial conditions of perturbations and are fully determined by the parameters of the compact object \cite{Kokkotas:1999bd,Berti:2009kk,Konoplya:2011qq}. The real part of the quasinormal mode represents a real oscillation frequency, while the imaginary one is proportional to the damping rate. Quasinormal modes of various wormholes were studied in a great number of publications \cite{Konoplya:2010kv,Bronnikov:2012ch,Taylor:2014vsa,Cuyubamba:2018jdl,Volkel:2018hwb,Aneesh:2018hlp,Konoplya:2018ala,Kim:2018ang,Roy:2019yrr,Churilova:2019qph,Jusufi:2020mmy}. Therefore, here we will concentrate on those 
examples of wormholes which include the parametric transition between the black hole and wormhole states. This way we will be able to understand whether the suggested parametrization provides a reasonable and compact approximation near the threshold of transition, that is, 
whether it allows one to describe wormholes as possible black hole mimickers \cite{Damour:2007ap,Cardoso:2016rao,Konoplya:2016hmd,Churilova:2019cyt,Bronnikov:2019sbx}. 
The general covariant equation for an electromagnetic field has the form
\begin{equation}\label{EmagEq}
\frac{1}{\sqrt{-g}}\partial_\mu \left(F_{\rho\sigma}g^{\rho \nu}g^{\sigma \mu}\sqrt{-g}\right)=0\,,
\end{equation}
where $F_{\rho\sigma}=\partial_\rho A_{\sigma}-\partial_\sigma A_{\rho}$,
and $A_\mu$ is a vector potential.
After separation of the variables Eq. (\ref{EmagEq}) for the spherically symmetric static spacetime  (Eq.~\eqref{le_general_shadows})  takes the following general wave-like form
\begin{equation}\label{wave-equation}
\frac{d^2\Psi}{dr_*^2}+\left(\omega^2-V(r)\right)\Psi=0\,,
\end{equation}
where the ``tortoise coordinate'' $r_*$ is defined in terms of the general metric functions $f(r)$ and $h(r)$ of Eq.~\eqref{le_general_shadows} by the relation
\begin{equation}
dr_*=\frac{dr}{\sqrt{f(r) h(r)}}\,,
\end{equation}
and the effective potential is 
\begin{eqnarray}\label{empotential}
V_{em}(r) = f(r)\frac{\ell(\ell+1)}{r^2}\,.
\end{eqnarray}
Notice that although the effective potential depends only on one of the two metric functions, the quasinormal modes will depend on both because the tortoise coordinate includes both. 

The effective potential for wormholes may have different forms. For example, it can have a single maximum located at the wormhole throat, or a couple of peaks, which are situated symmetrically relative to the throat. A common feature is that at ``minus infinity'' and
``plus infinity'' the effective potential  approaches some constant values representing the 
asymptotic flatness of the spacetime. 
Therefore the boundary conditions for finding quasinormal modes of a wormhole are similar 
to those for a black hole in the tortoise coordinate \cite{Konoplya:2005et}: purely 
outgoing waves are required at both infinities.

Here we shall use the time-domain integration method, which does not depend on the form 
of a potential barrier and can be applied to all cases under consideration. We will integrate the wave-like equation rewritten in terms of the light-cone variables $u=t-r_*$ and $v=t+r_*$ 
with the help of the Gundlach-Price-Pullin discretization scheme \cite{Gundlach:1993tp}:
\begin{eqnarray}\label{Discretization}
\Psi\left(N\right)&=&\Psi\left(W\right)+\Psi\left(E\right)-\Psi\left(S\right)\nonumber
\\	
&&-\Delta^2\frac{V\left(W\right)\Psi\left(W\right)+V\left(E\right)\Psi\left(E\right)}{8}	+{\cal O}\left(\Delta^4\right)\,,
\end{eqnarray}
where we used the following designations for the points:
$N=\left(u+\Delta,v+\Delta\right)$, $W=\left(u+\Delta,v\right)$, $E=\left(u,v+\Delta\right)$, and  
$S=\left(u,v\right)$. The initial data are given on the null surfaces $u=u_0$ and $v=v_0$.  
To extract the values of the quasinormal frequencies, we will use the Prony method 
which allows one to fit the signal by a sum of exponents with some excitation factors. 
This method was used in a great number of works (see, for instance, the recent papers \cite{Churilova:2020bql,Konoplya:2020bxa,Churilova:2020aca,Konoplya:2020jgt} and references therein) and showed a good agreement with accurate results obtained by other methods.

Quasinormal modes of the Bronnikov-Kim braneworld wormhole with zero Schwarzschild mass, the CFM wormhole and the Simpson-Visser metrics are presented in Tables \ref{QNM1}, \ref{QNM2}, \ref{QNM3}.

\begin{table}[H]
\center{
\caption{Fundamental quasinormal modes ($\ell=1$, $n=0$)  of the electromagnetic field for Bronnikov-Kim II wormhole.}
\begin{tabular}{|c|c|c|c|}
 \hline
 order
  & $C=-0.01$  & $C=-0.3$ &  $C=-0.7$ \\
 \hline\hline
  exact & $0.64306 - 0.20529 i$ & $0.70322 - 0.03991 i$ & $0.73338 - 0.09961 i$ \\
 \hline
   $1$st order & $0.65899-0.19306 i$  & $0.70232-0.03741 i$ & $0.73304-0.09836 i$ \\
 \hline
   $2$nd order & $ 0.64093-0.20593 i$  & $0.70332-0.04009 i$ & $0.73345-0.09969 i$   \\
 \hline
\end{tabular}\label{QNM1}
}
\end{table}

\begin{table}[H]
\center{
\caption{Fundamental quasinormal modes ($\ell=1$, $n=0$)  of the electromagnetic field for CFM wormhole ($m=1/2$).}
\begin{tabular}{|c|c|c|c|}
 \hline
 order
  & $r_{0}=2.2\,m$  & $r_{0}=20\,m$ &  $r_{0}=40\,m$ \\
 \hline\hline
  exact & $0.50177 - 0.11188 i$ & $0.12759 - 0.03054 i$ & $0.06552 - 0.01498 i$ \\
 \hline
   $1$st order & $0.50261 - 0.11025 i$  & $0.12757 - 0.03053 i$  & $0.06553 - 0.01497 i$ \\
 \hline
\end{tabular}\label{QNM2}
}
\end{table}

\begin{table}[H]
\center{
\caption{Fundamental quasinormal modes ($\ell=1$, $n=0$)  of the electromagnetic field for Simpson-Visser wormhole ($m=1/2$).}
\begin{tabular}{|c|c|c|c|}
 \hline
 order
  & $a = 1.01$  & $a=1.25$ &  $a = 1.4$ \\
 \hline\hline
  exact & $0.51125 - 0.13311 i$ & $0.55491 - 0.03299 i$ & $0.56858 - 0.05986 i$ \\
 \hline
   $1$st order & $0.50899 - 0.13211 i$  & $0.55394 - 0.03297 i$  & $0.56770 - 0.05982 i$ \\
 \hline
   $2$nd order & $0.51241 - 0.13405 i$  & $0.55520 - 0.03286 i$  & $0.56885 - 0.05969 i$   \\
 \hline
\end{tabular}\label{QNM3}
}
\end{table}

The Prony method does not allow for extracting quasinormal frequencies with guaranteed accuracy higher than a fraction of one percent. This happens because the time at which the beginning and end of quasinormal ringing process occurs is conditional. Therefore, when using time-domain integration, there is no point in looking at high orders of the parametrization, because the error of the Prony method will be evidently larger than that of the parametrization. Nevertheless, this method allows us to judge about the convergence of the parametrization at the first few orders. From the above tables we see that far from the threshold of the wormhole/black hole transition, already the first order of expansion makes a relative error about one percent or less. At the same time, the period of quasinormal ringing is very short near the transition and is ``contaminated'' by consequent echoes, as can be seen, for example, from Fig.\,2 in \cite{Bronnikov:2019sbx}. Therefore, we are unable to see whether the first-order expansion is really bad near the transition point, but we see that already at the second order (for the Bronnikov-Kim II metric) the situation is remedied.  

\vspace{0.5cm}
%===============================================
\section{Conclusions \label{Sec:conclusions}}
%===============================================

\noindent In this work we have developed a general parametrization for static, spherically symmetric and asymptotically flat wormhole geometries in an arbitrary metric theory of gravity. We have also extended the parametrization to a slow rotation mode. We used the previously developed general parametrization of black holes \cite{Rezzolla:2014mua}, which we adopted 
for wormholes by a number of appropriate modifications. The continued-fraction expansion in the radial direction in terms of a compact coordinate provides superior convergence, and the various examples of analytic wormhole metrics, which we considered for testing our parametrization, have shown that a very good accuracy is achieved already at the first order 
of the expansion, while the second order guarantees that the relative error is less than one percent in all considered cases. 
Remarkably, the method is convergent also at a transition between the black hole and wormhole states, so that wormholes mimicking black holes' behavior can also be well described by the above parametrization. The latter is confirmed by our calculations of the radii of shadows cast by the wormholes and by their quasinormal spectra.

In the present paper we restricted ourselves to analytical wormhole metrics, because, to the best of our knowledge, no numerical solution is known for a four-dimensional, traversable, asymptotically flat wormhole, which would not require exotic matter and be stable against small perturbations of spacetime.
By now there are two numerical wormhole solutions  \cite{Kanti:2011jz,Blazquez-Salcedo:2020czn} constructed without adding exotic matter. The first solution \cite{Kanti:2011jz} was found in the Einstein-dilaton-Gauss-Bonnet theory and was later extended to various couplings of the scalar field in \cite{Antoniou:2019awm}. In \cite{Cuyubamba:2018jdl} it was shown that the wormhole with exponential (dilatonic) coupling is unstable. Even though in the regime of small coupling of a scalar field, the dilatonic term is dominating, and thereby the instability is highly expected, a stability analysis in the case of other, nondilatonic, forms of the scalar field has not been performed so far. The second solution \cite{Blazquez-Salcedo:2020czn} has been recently found when adding coupled Dirac and Maxwell fields to the gravitational one, and its stability has not been tested as well. If in the course of studying scalar and gravitational perturbations of these wormholes, some of them appear to be stable, analytical approximations for such metrics could be readily found with the help of the present formalism, and then their observational properties can be analyzed along the same lines as is done in this paper.

Our paper could be further extended to arbitrary rotation in a similar fashion with the black hole case \cite{Konoplya:2016jvv}, that is, by expansion in terms of $\cos \theta$ around the equatorial plane. However, apart from the evident lack of a sufficient number of examples in the literature, apparently this most general parametrization of axially symmetric wormholes  will not share the elegance and simplicity of interpretation of the case considered here, and therefore deserves a separate consideration.

\begin{acknowledgments}
R. K. and T. P. acknowledge the support of the grant 19-03950S of Czech Science Foundation (GA\v{C}R). K. B. was supported by the RUDN University Strategic Academic Leadership Program. The research of K. B. was also funded by the Ministry of Science and Higher Education of the Russian Federation, Project ``Fundamental properties of elementary particles and cosmology'' N 0723-2020-0041, and by RFBR Project 19-02-00346.
\end{acknowledgments}

%\newpage
\vspace{1 cm}
\bibliography{References}{}
\bibliographystyle{utphys}
\end{document}